\documentclass[a4paper,twocolumn]{esapub} 

\def\figdir{FIGS-PS}
\usepackage{times}
\usepackage{graphicx}
\usepackage{natbib}
\usepackage{journals}

\def\corot{CoRoT}
\def\astec{{\small\bf ASTEC}}
\def\cesam{{\small\bf CESAM}}
\def\cles{{\small\bf CL\'ES}}
\def\franec{{\small\bf FRANEC}}
\def\genec{{\small\bf GENEC}}
\def\starox{{\small\bf STAROX}}
\def\tgec{{\small\bf TGEC}}
\def\adipls{{\small\bf ADIPLS}}
\def\filou{{\small\bf FILOU}}
\def\graco{{\small\bf GRACO}}
\def\losc{{\small\bf LOSC}}
\def\noc{{\small\bf NOC}}
\def\oscrox{{\small\bf OSCROX}}
\def\posc{{\small\bf POSC}}
\def\romosc{{\small\bf ROMOSC}}
\def\elem#1[#2]{{}^{#2}{\rm #1}}

\begin{document}
\title{Report on the C{\small o}R{\small o}T Evolution and Seismic Tools Activity}
\author[1,2]{M.J.P.F.G.~Monteiro}
\author[3]{Y.~Lebreton}
\author[4]{J.~Montalb\'an}
\author[5]{J.~Christensen-Dalsgaard}

\author[6]{M.~Castro}
\author[7]{S.~Degl'Innocenti}
\author[8]{A.~Moya}
\author[9,8]{I.W.~Roxburgh}
\author[4]{R.~Scuflaire}

\author[8]{A.~Baglin}
\author[1]{M.S.~Cunha}
\author[10]{P.~Eggenberger}
\author[11]{J.~Fernandes}
\author[3]{M.J.~Goupil}
\author[6]{A.~Hui-Bon-Hoa}
\author[12]{M.~Marconi}
\author[1,2,11]{J.P.~Marques}
\author[3]{E.~Michel}
\author[4]{A.~Miglio}
\author[13]{P.~Morel}
\author[13]{B.~Pichon}
\author[7]{P.G.~Prada~Moroni}
\author[13]{J.~Provost}
\author[12]{A.~Ruoppo}
\author[14,3]{J.-C.~Su\'arez}
\author[15]{M.~Suran}
\author[1]{T.C.~Teixeira}

\affil[1]{Centro de Astrof\'{\i}sica da Universidade do Porto, Portugal}
\affil[2]{Departamento de Matem\'atica Aplicada, Faculdade de Ci\^encias da Universidade do Porto, Portugal}
\affil[3]{Observatoire de Paris, GEPI, CNRS UMR 8111, Meudon, France}
\affil[4]{Institut d'Astrophysique et de G\'eophysique de l'Universit\'e de Li\`ege, Belgium}
\affil[5]{Institut for Fysik og Astronomi, Aarhus Universitet, Denmark}
\affil[6]{LATT Observatoire Midi-Pyr\'en\'ees, CNRS UMR 5572, Universit\'e Paul Sabatier, Toulouse, France}
\affil[7]{Dipartimento di Fisica, Universit\'a di Pisa, Italy}
\affil[8]{Observatoire de Paris, LESIA, CNRS UMR 8109, Meudon, France}
\affil[9]{Queen Mary University of London, England}
\affil[10]{Observatoire de Gen\`eve, Switzerland}
\affil[11]{Grupo de Astrof\'{\i}sica, Observat\'orio Astron\'omico da Universidade de Coimbra, Portugal}
\affil[12]{INAF-Osservatorio Astronomico di Capodimonte, Napoli, Italy}
\affil[13]{D\'epartement Cassiop\'ee, CNRS UMR 6202, Observatoire de la C\^ote d'Azur, Nice, France}
\affil[14]{Instituto de Astrof\'{\i}sica de Andaluc\'{\i}a (CSIC), Granada, Spain}
\affil[15]{Astronomical Institute of the Romanian Academy, Romania}


\maketitle
\begin{abstract}

  We present the work undertaken by the \textit{Evolution and Seismic Tools Activity} (ESTA) team of the \corot\ \textit{Seismology Working Group}.
  We have focused on two main tasks: Task~1 -- now finished -- has aimed at testing, comparing and optimising seven stellar evolution codes which will be used to model the internal structure and evolution of the \corot\ target stars.
  Task~2, still underway, aims at testing, comparing and optimising different seismic codes used to calculate the oscillations of  models for different types of stars.
  The results already obtained are quite satisfactory, showing minor differences between the different numerical tools provided the same assumptions on the physical parameters are made.
  This work gives us confidence on the numerical tools that will be available to interpret the future \corot\ seismic data.
\end{abstract}

\section{Introduction}\label{sec:intro}

  The space experiment \corot\ will provide us with stellar oscillation data (frequencies, amplitudes, line widths) for stars of various masses and chemical compositions -- mainly main sequence solar type stars and $\delta$ Scuti stars -- with an expected accuracy on the frequencies of ${\sim}0.1~\mu$Hz \citep{2002sshp.conf...17B}.
  Such an accuracy is needed for the determination of the position of the external convection zone in solar type stars \citep{1994A&A...282...73A,monteiro00,monteiro02} or the extent of the convective core of intermediate mass stars \citep{1999ASPC..173..257R} as well as for the inversion of the stellar density profile \citep{1993MNRAS.264..522G} and of the rotation profile or for the separation of multiplets \citep{1996A&A...305..487G}.
  However, studies in helioseismology have taught us that in order to probe fine details of the solar internal structure, we need both the constraints of high quality seismic data and very accurate solar models \citep{1982MNRAS.199..735C,1995MNRAS.274..899R,gough96,monteiro02}.
  Therefore, to be able to draw valuable information from the future \corot\ data, we have to ensure now that we will be able to interpret them with models having reached the required level of accuracy.

  Within the \corot\ Seismology Working Group, the ESTA group has been set up, with the aim to extensively test, compare and optimise the numerical tools used to calculate stellar models and their oscillation frequencies.
  Our goals are (i) to be able to produce theoretical seismic predictions by means of different numerical codes and to understand the possible differences between them and (ii) to bring stellar models at the level of accuracy required to interpret the future \corot\ seismic data.

  Up to now, the ESTA group has focused on specific tasks.
  The first one -- Task~1 -- has consisted in comparing stellar models and evolution sequences produced by seven participating evolution codes, presented in Section~\ref{sec:tools}.
  In this context we have calculated models for several specific study cases, presented in Section~\ref{sec:task1}, corresponding to stars covering the range of masses, evolution stages, and chemical compositions expected for the bulk of \corot\ target stars.
  The main comparisons made for Task~1 are presented and discussed in Section~\ref{sec:task1-results}.
  The second -- Task~2 -- is still in progress.
  It consists of testing, comparing and optimising the seismic codes by comparing the frequencies produced by different oscillation codes, also presented in Section~\ref{sec:tools}, again for specific stellar cases.
  The preliminary results of Task~2 are presented in Section~\ref{sec:task2}.
  The \corot\ stellar model grids are described in Section~\ref{sec:grids}.
  In Section~\ref{conclusion} we present our conclusions and plans for future studies.
  
  All the results presented here are a synthesis of the work presented during four ESTA meetings held in 2005, in Toulouse-France (May, $8^{\mbox{\tiny th}}$ \corot\ Week -- CW8), in Nice-France (September), in Aarhus-Denmark (October) and at ESA-ESTEC (December, CW9).
  The corresponding presentations and posters can be consulted at the ESTA Web site at {\tt\small http://www.astro.up.pt/corot/}.

\section{Numerical tools for ESTA}\label{sec:tools}

\subsection{Stellar internal structure and evolution codes}

  Seven stellar evolution codes have been used to calculate the models to be compared.
  We give here brief information on where to find documentation about these codes.

\begin{itemize}

\item[(A)] \astec\ -- {\em Aarhus Stellar Evolution Code}:
  a general description of the original code and updates can be found in \citet{1982MNRAS.199..735C,CD05ab}.
  
\item[(C)] \cesam\ -- {\em Code d'\'Evolution Stellaire Adaptatif et Modulaire}:  
  the original code and updates are described in \citet{Morel97} and \citet{PichonMorel05} and at the Website:\\
  {\tt\small http://www.obs-nice.fr/cesam/}.
  
\item[(L)] \cles\ -- {\em Code Li\`egeois d'\'Evolution Stellaire}:
  a description of the code, which is still in an active phase of development, can be found in \citet{Scuflaire05}.
  
\item[(F)] \franec\ -- {\em Pisa Evolution Code}:
  information on the code and its updates can be found in
\citet{cariulo04} and \citet{DeglInn05}.

\item[(G)] \genec\ -- {\em Geneva Evolution Code}:
  the main properties and physical assumptions of the code are discussed in \citet{meynet00}.

\item[(S)] \starox\ -- {\em Roxburgh's Evolution Code}:
  a description of the code can be found in \citet{roxburgh05a,roxburgh05b}.
  
\item[(T)] \tgec\ -- {\em Toulouse-Geneva Evolution Code}:
  information on the code can be found in
\citet{1996A&A...312.1000R} and \citet{Castro05}.

\end{itemize}

\subsection{Stellar oscillation codes}

  Eight stellar oscillation codes are available to us to calculate the oscillation frequencies of models.

\begin{itemize}

\item \adipls\ -- {\em Aarhus Adiabatic Pulsation Package}:
  available at:\\
  {\tt\small http://astro.phys.au.dk/$\sim$jcd/adipack.n}

\item \filou\ -- {\em Meudon Oscillations Code}:
  see \citet{2002PhDT.........6S}.

\item \graco\ -- {\em Granada Oscillation Code}:
  see \citet{2004A&A...414.1081M}.

\item \losc\ -- {\em Li\`ege Oscillations Code}:
  see \citet{1975A&A....41..279B,Scuflaire05}. 

\item \noc\ -- {\em Nice Oscillations Code}:
  see \citet{1989nos..book.....U}.

\item \oscrox\ -- {\em Roxburgh's Oscillations Code}:
  see \citet{roxburgh05b}.

\item \posc\ -- {\em Porto Linear Adiabatic Oscillations Code}:
  see: {\tt\small http://www.astro.up.pt/$\sim$mjm/}.

\item \romosc\ -- {\em Linear Non-Adiabatic Non-Radial Waves}:
  see \citet{suran91}.
  
\end{itemize}

\section{Task~1: Inputs for model comparison}\label{sec:task1}

\subsection{Input physics for the codes}\label{sec:physics}

  All the models to be compared have been calculated with a given set of standard input physics.
  In this first step we have neglected microscopic and turbulent diffusion processes as well as rotation and magnetic fields.

\begin{itemize}

\item \underbar{Equation of State}:
  We used the OPAL2001 \citep{2002ApJ...576.1064R} equation of state which is provided in a tabular form.
  In most codes all the thermodynamic quantities are obtained from the variables $\rho$, $T$, $X$ and $Z$ (respectively density, temperature, hydrogen and heavy element mass fraction) using the interpolation package provided on the OPAL Web site.
  But in some codes (\cles, \starox) a set of thermodynamical quantities are interpolated in the OPAL tables  ($C_{V}$, $P$, $\chi_\rho$ and $\chi_{T}$ for \cles) by a method ensuring the continuity of first derivatives at cell boundaries in the 4D space defined by the variables $\rho$, $T$, $X$ and $Z$.
  The other thermodynamic quantities (in \cles\ $\Gamma_1$, $\Gamma_{3}{-}1$ and $C_{P}$) are derived from the values of $C_{V}$, $P$, $\chi_\rho$ and $\chi_{\rm T}$ using the thermodynamic relations.

\item \underbar{Opacities}:
  We used the OPAL95 opacity tables \citep{ir96} complemented at low temperatures by the \citet{af94} tables.
  The interpolation methods used by the codes may differ.
  In all codes the metal mixture of the opacity tables is fixed within a given model.

\item \underbar{Nuclear reaction rates}:
  We used the basic pp and CNO reaction networks up to the $\elem{O}[17](p,\alpha)\elem{N}[14]$ reaction.
  Depending of the code the combustion of $\elem{Li}[7]$ and $\elem{H}[2]$ is entirely followed (\cles, \franec) or these elements are assumed to be at equilibrium (\astec, \cesam, \starox).
  The nuclear reaction rates are computed using the analytical formulae provided by the NACRE compilation \citep{1999NuPhA.656....3A}.
  Weak screening is assumed under \citet{1954AuJPh...7..373S}'s formulation.

\item \underbar{Convection and overshooting}:
  We use the classical mixing length treatment of  \citet{1958ZA.....46..108B} under the formulation of \citet{1965ApJ...142..841H} taking into account the optical thickness of the convective bubble.
  The onset of convection is determined according to the Schwarzschild criterion $(\nabla_{\rm ad}{-}\nabla_{\rm rad}{<}0)$ where $\nabla_{\rm ad}$ and $\nabla_{\rm rad}$ are respectively the adiabatic and radiative temperature gradient.
  We adopt a  mixing-length parameter $\alpha_{\mbox{\scriptsize{MLT}}}{=}1.6$.
  In models with overshooting, the convective core is extended on a distance $\ell_{\mbox{\scriptsize ov}}{=}\alpha_{\mbox{\scriptsize ov}}{\times}\min(H_{P}, R_{\mbox{\small cc}})$ where $\alpha_{\mbox{\scriptsize ov}}{=}0.15$ is the chosen overshooting parameter, $H_{P}$ is the pressure scale height and $R_{\mbox{\small cc}}$ is the radius of the convective core. 
  The core is fully mixed in the region corresponding to the convective and overshooting region; in the overshooting region the temperature gradient is taken to be equal to the adiabatic gradient.  

\item \underbar{Atmosphere}:
  Eddington's grey $T(\tau)$ law is used for the atmosphere calculation: $T{=}T_{\rm eff}[\frac{3}{4}(\tau{+}\frac{2}{3})]^{1/4}$ where $\tau$ is the optical depth.
  The level where the integration of the hydrostatic equation starts depends on the codes as well as the level where the atmosphere is matched to the envelope.
  The radius of the star is taken to be the bolometric radius, i.e. the radius at the level where the local temperature equals the effective temperature ($\tau{=}2/3$ for the Eddington's law).

\item \underbar{Initial abundances} of the elements and heavy elements mixture:
  All models are calculated with the classical \citet[][hereafter GN93]{gn93} solar mixture of heavy elements.
  The mass fractions of hydrogen ($X$), helium ($Y$) and heavy elements ($Z$) are specified for each model.
  In the nuclear reaction network the initial abundance of each chemical species is split between its isotopes according to the isotopic ratio of nuclides.

\item \underbar{Physical and astrophysical constants}:
  We used the reference values listed at \citet{monteiro05b}.

\end{itemize}

\begin{table}[ht!]
\begin{center}
\caption{
  Target models for Task~1 (see Fig.~\ref{fig:hr_targets}).
  Masses ($M$ and $M^{\mbox{\scriptsize He}}_{\mbox{\scriptsize c}}$) are given in units of solar mass ($M_\odot$) while temperature is in K.
  ``OV'' indicates that overshoot has been included (see text).}
\vspace{5pt}  
\begin{tabular}[h]{cccclr}
\hline\\[-8pt]
{\bf Case} & \boldmath$M$ & \boldmath$X_0$ & \boldmath$Z_0$ &
  {\bf Specification} & {\bf Type} \\
\hline\\[-8pt]
{\bf 1.1} & 0.9 & 0.70 & 0.02 &
  $X_{\mbox{\scriptsize c}}{=}0.35$ & 
  \begin{small}MS\end{small}\\[2pt]
{\bf 1.2} & 1.2 & 0.70 & 0.02 &
  $X_{\mbox{\scriptsize c}}{=}0.69$ & 
  \begin{small}ZAMS\end{small}\\[2pt]
{\bf 1.3} & 1.2 & 0.73 & 0.01 &
  $M^{\mbox{\scriptsize He}}_{\mbox{\scriptsize c}}{=}0.1$ & 
  \begin{small}SG\end{small}\\[2pt]
{\bf 1.4} & 2.0 & 0.70 & 0.02 &
  $T_{\mbox{\scriptsize c}}{=}1.9{\times}10^7$ &
  \begin{small}PMS\end{small}\\[2pt]
{\bf 1.5} & 2.0 & 0.72 & 0.02 &
  ${\mbox{OV}}, X_{\mbox{\scriptsize c}}{=}0.01$ & 
  \begin{small}TAMS\end{small}\\[2pt]
{\bf 1.6} & 3.0 & 0.71 & 0.01 &
  $X_{\mbox{\scriptsize c}}{=}0.69$ &
  \begin{small}ZAMS\end{small}\\[2pt]
{\bf 1.7} & 5.0 & 0.70 & 0.02 &
  $X_{\mbox{\scriptsize c}}{=}0.35$ &
  \begin{small}MS\end{small}\\[2pt]
\hline
\end{tabular}
\label{tab:modspec}
\end{center}
\end{table}

\begin{figure}[ht!]
\centering
\hspace{-8pt}\includegraphics[width=0.9\linewidth]{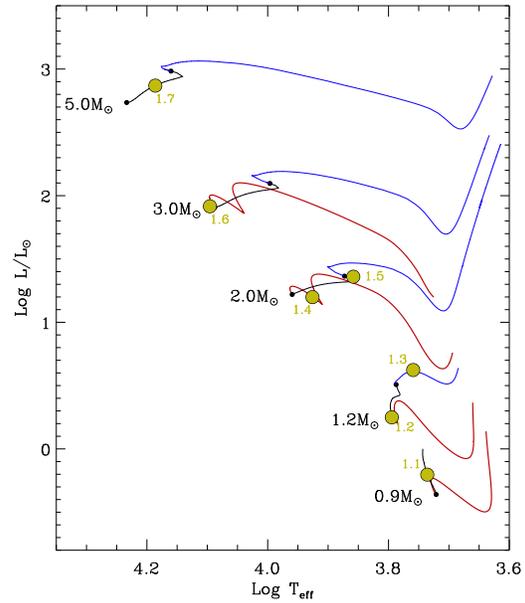}
\caption{
  HR diagram showing the targets for Task~1 (see Table~\ref{tab:modspec}).
  Red lines correspond to the PMS, black lines to the MS and blue lines to the post MS evolution (\corot\ reference grid from \cesam).
  The targets are ordered in mass and age along the diagram, from Case 1.1 (bottom-right) up to Case 1.7 (top-left).}
\label{fig:hr_targets}
\end{figure}

\subsection{Initial parameters of the models}

\begin{figure}[ht!]
\centering
\begin{tabular}{cccc}
\hspace{-8pt}\includegraphics[width=0.9\linewidth]{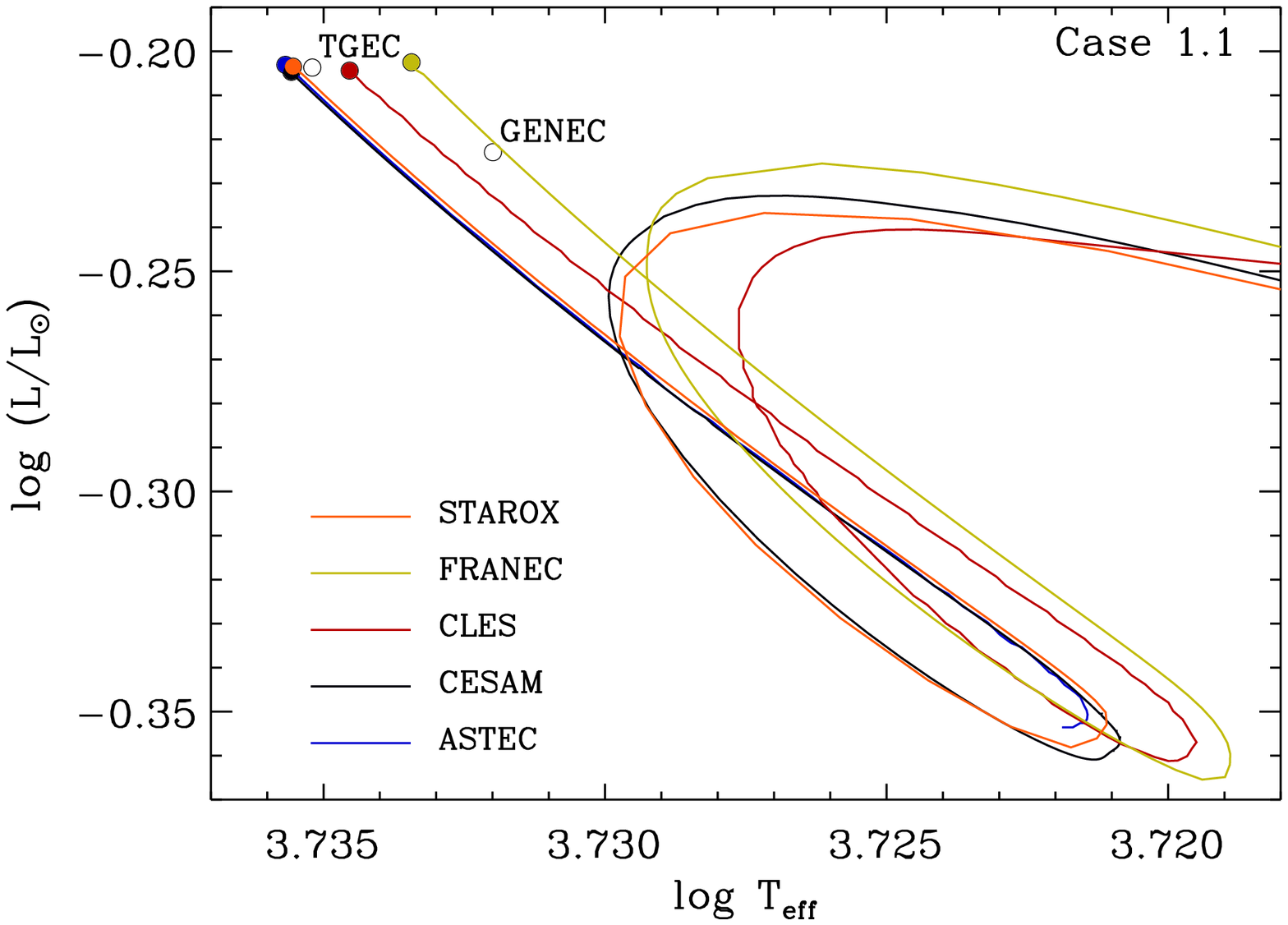}\\[-5pt]
\hspace{-8pt}\includegraphics[width=0.9\linewidth]{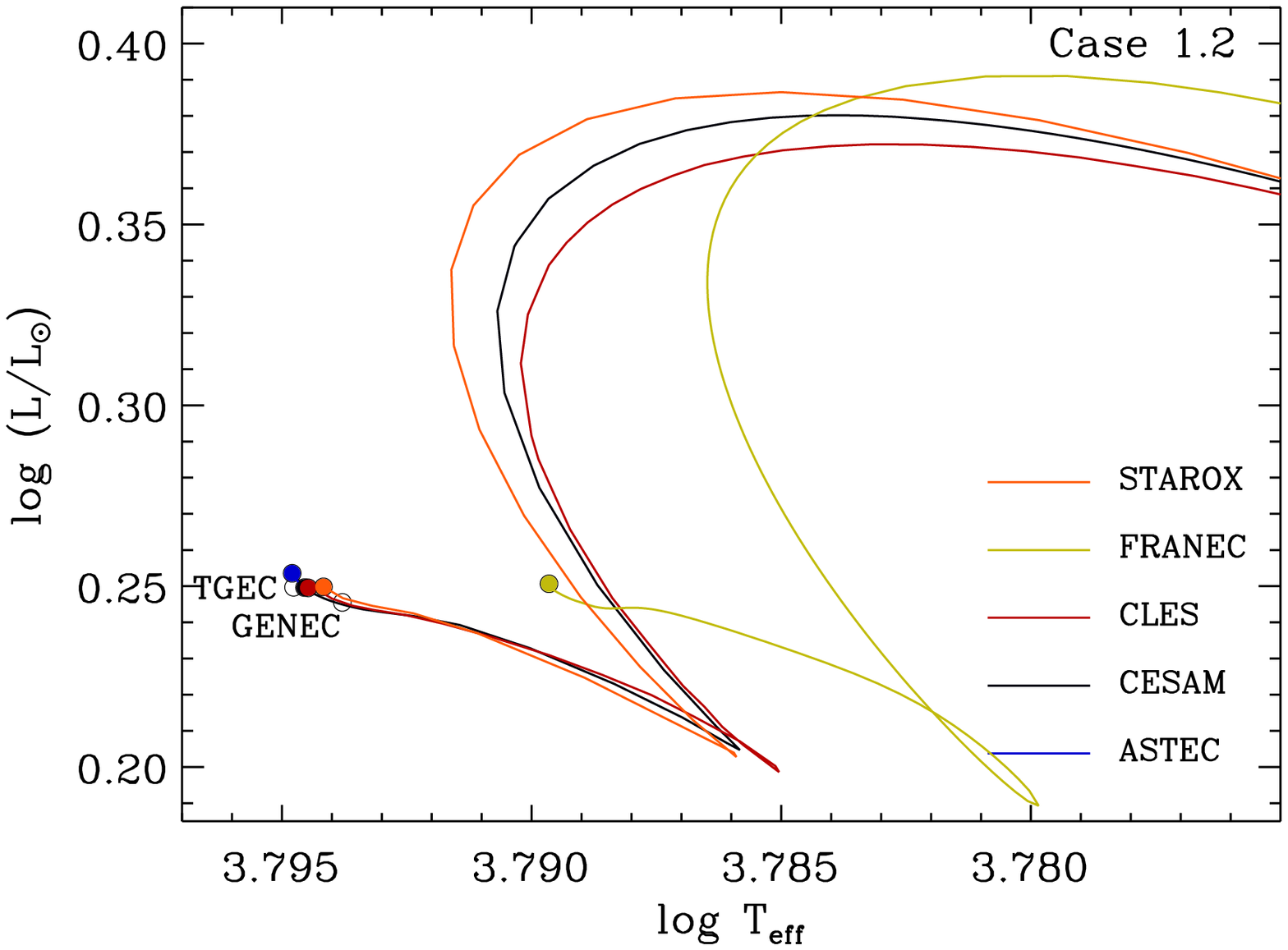}\\[-5pt]
\hspace{-8pt}\includegraphics[width=0.9\linewidth]{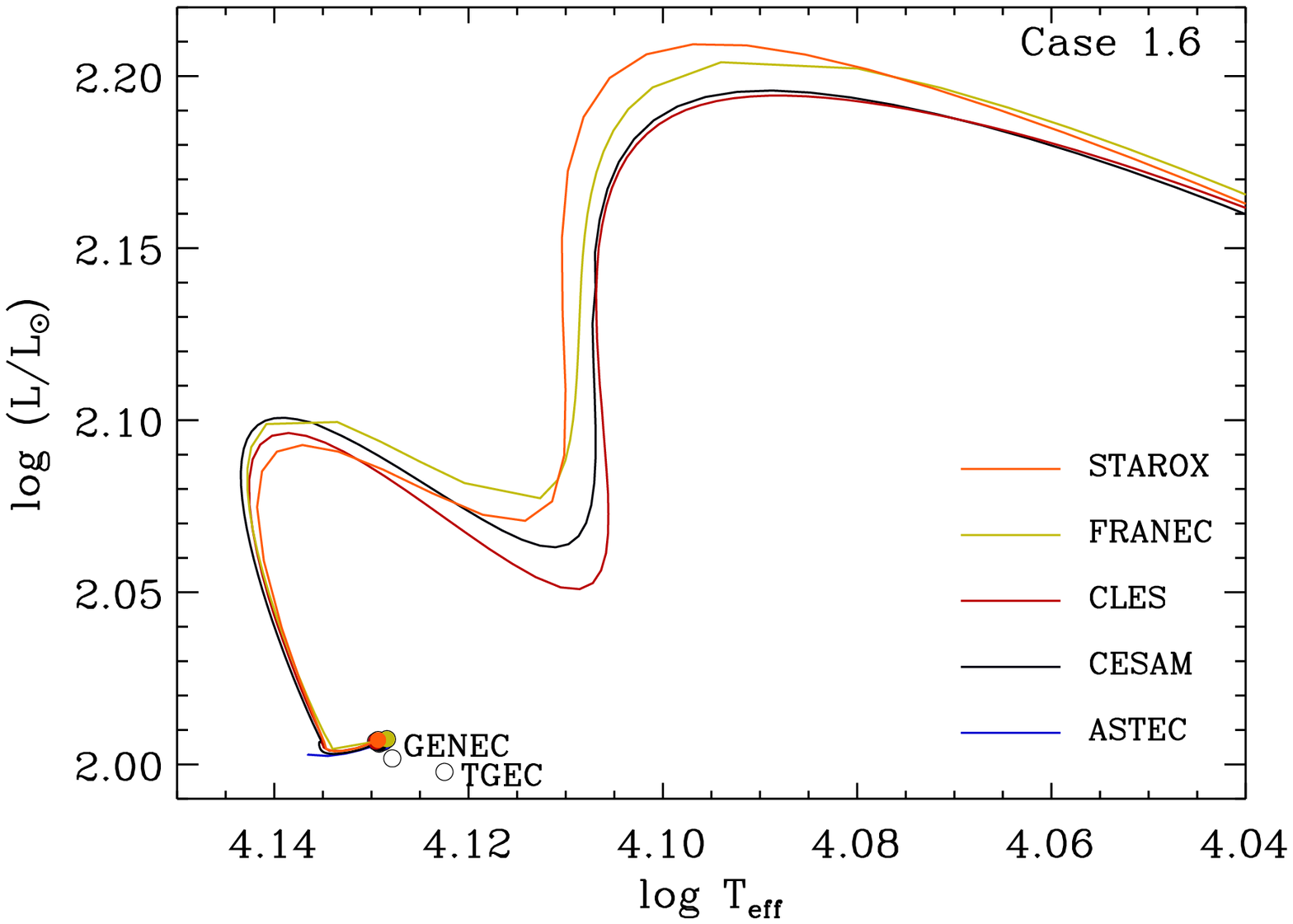}\\[-5pt]
\hspace{-8pt}\includegraphics[width=0.9\linewidth]{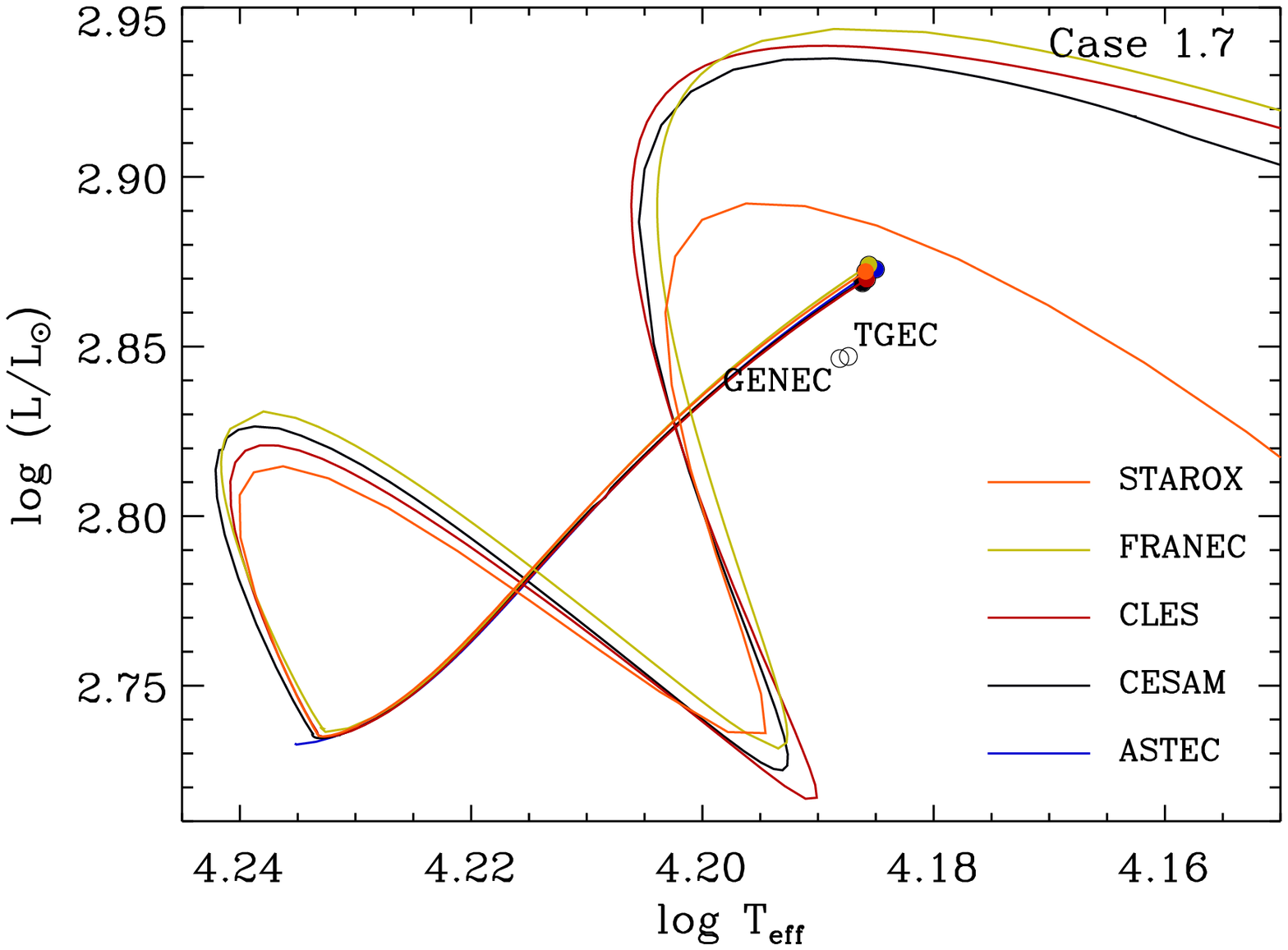}
\end{tabular}
\caption{
  Evolutionary sequences and final target model for Cases 1.1, 1.2, 1.6 and 1.7 (all are MS).}
\label{fig:dif_hr_inmsa}
\end{figure}

  We have focused on a few cases representative of the range of stellar masses, chemical composition and evolutionary stage of \corot's main targets. 
  Table~\ref{tab:modspec} presents the input parameters of the models while Figure~\ref{fig:hr_targets} gives their position in the HR diagram.
  For each model we examine both the evolution with time of the global parameters (luminosity, effective temperature, mass of the convective core) and the internal structure parameters at given evolution stages.
  We consider several evolution stages: one pre-main sequence (PMS) stage with central temperature $T_{\rm c}{=}1.9{\times}10^7$~K, two stages close to the zero-age main sequence (ZAMS), two main sequence (MS) stages having a central hydrogen content $X_{\rm c}$ of about half the initial value, a stage close to the core hydrogen exhaustion (TAMS), and a post MS subgiant stage (SG) in which $M^{\mbox{\scriptsize He}}_{\mbox{\small c}}$, the mass of the central region of the star where the hydrogen abundance is $X{\le}0.01$, is such that $M^{\mbox{\scriptsize He}}_{\mbox{\small c}}{=}0.10~M_\odot$. 
  
\begin{figure}[ht!]
\centering
\begin{tabular}{ccc}
\hspace{-8pt}\includegraphics[width=0.9\linewidth]{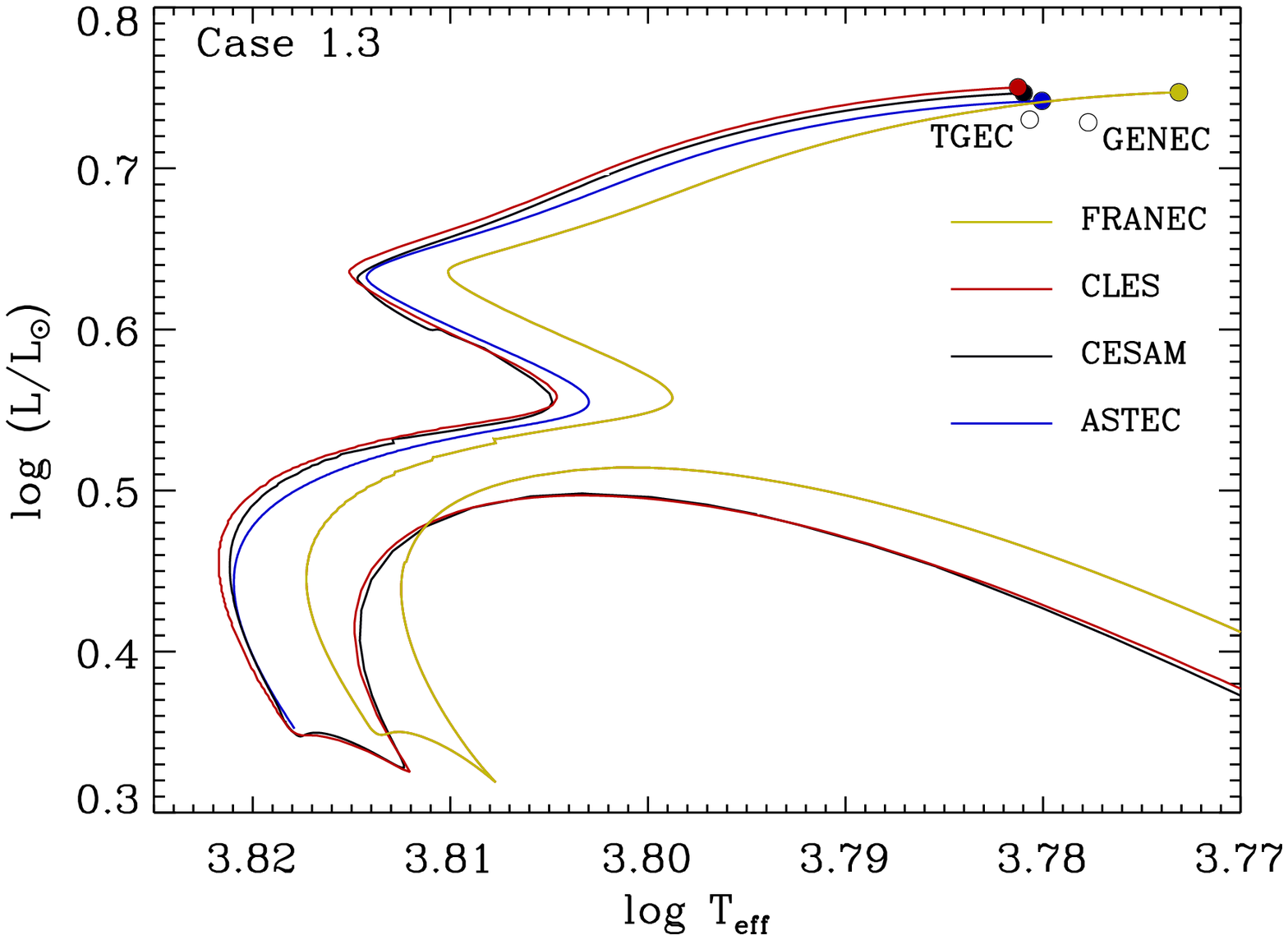}\\[-5pt]
\hspace{-8pt}\includegraphics[width=0.9\linewidth]{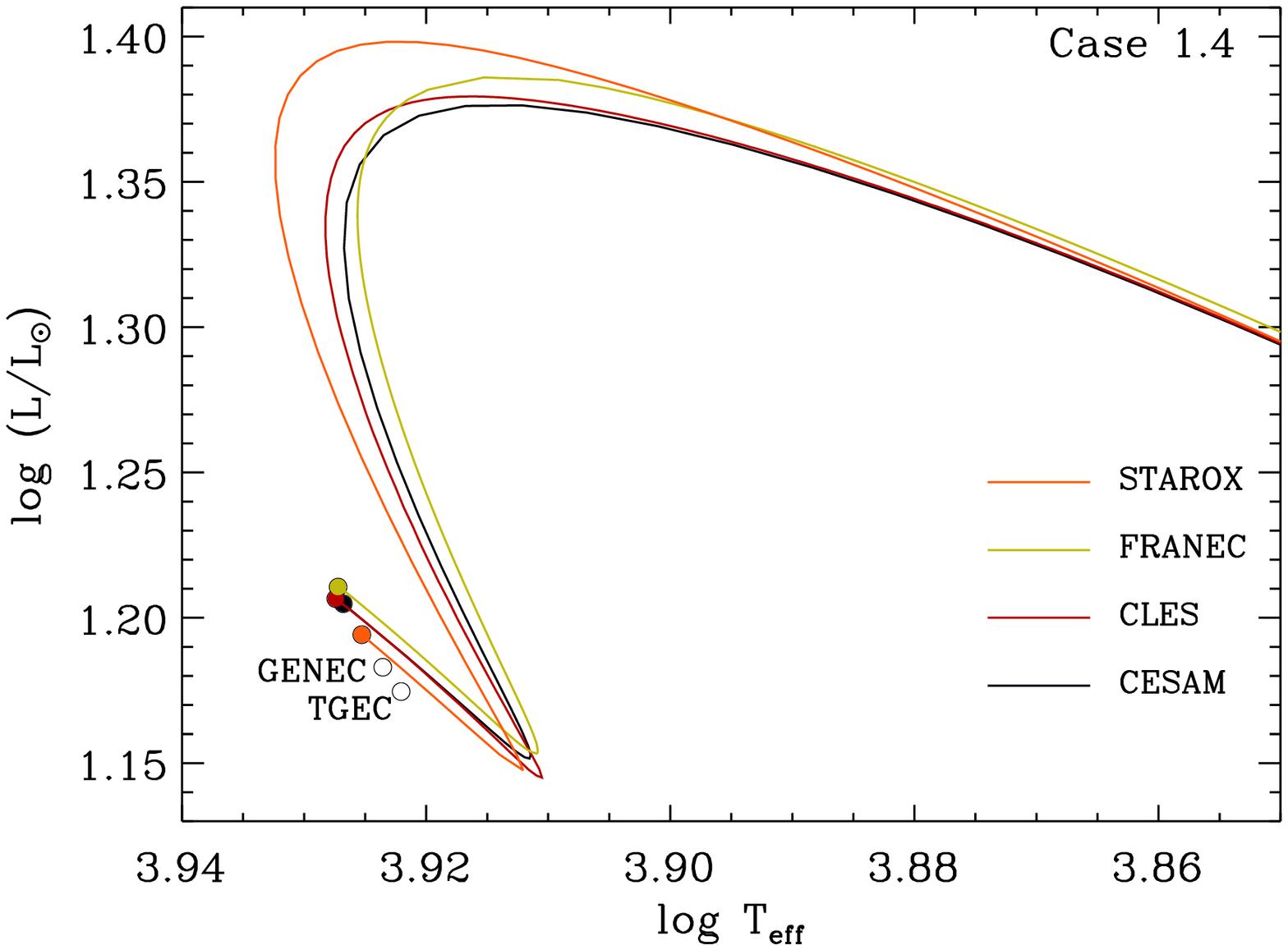}\\[-5pt]
\hspace{-8pt}\includegraphics[width=0.9\linewidth]{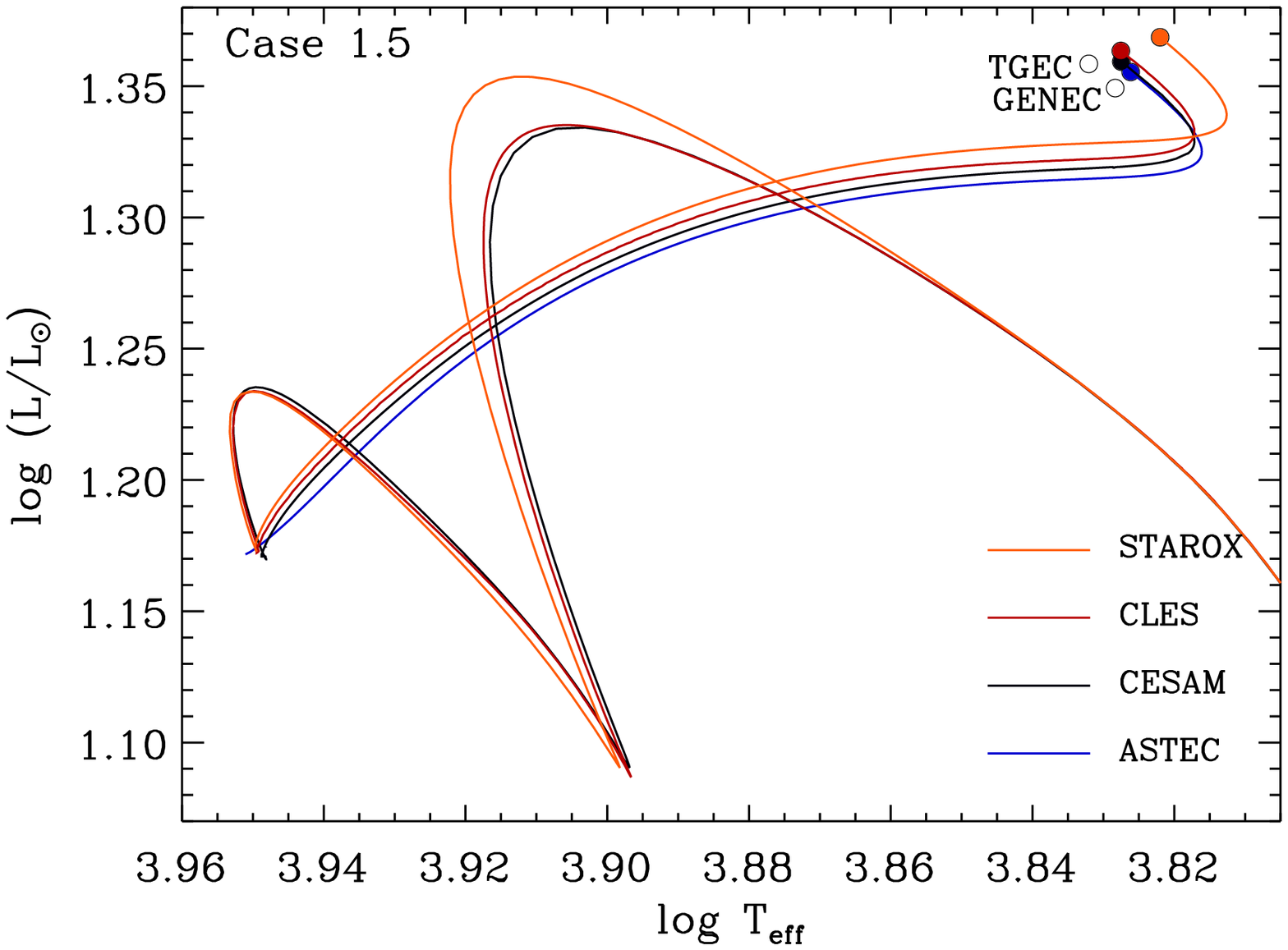}
\end{tabular}
\caption{
  Same as Figure~\ref{fig:dif_hr_inmsa} for Case 1.3 (SG), Case 1.4 (PMS) and Case 1.5 (TAMS).}
\label{fig:dif_hr_offms}
\end{figure}

\section{Task~1: Results of the comparison}\label{sec:task1-results}

\begin{table*}[ht!]
\begin{center}
\caption{
  Differences in the global parameters for all Cases (see Table~\ref{tab:modspec}) calculated using different stellar evolutionary codes.
  [all] are the results obtained when using the global parameters from all codes; [3] when using the parameters of models from codes: (A, C, L) in Case 1.3, (C, L, S) in Case 1.4; [4] is for differences between parameters of models from codes (A, C, L, S).}
\vspace{1em}             
\renewcommand{\arraystretch}{1.2}
\begin{tabular}{|l|rr|rr|rr|rr|rr|rr|}
\hline
 & \multicolumn{2}{c}{{\bf Age} (Myr)}
 & \multicolumn{2}{|c|}{\boldmath$R/R_\odot$}
 & \multicolumn{2}{c}{\boldmath$L/L_\odot$}
 & \multicolumn{2}{|c|}{{\boldmath$T_{\mbox{\bf\scriptsize eff}}$}~\mbox{(K)}}
 & \multicolumn{2}{c}{{\boldmath$T_{\mbox{\bf\scriptsize c}}$}~($10^7$K)}
 & \multicolumn{2}{|c|}{\boldmath$\rho_{\mbox{\bf\scriptsize c}}$~\mbox{(cgs)}}\\ 
\hfill {\bf Case}  & 
\boldmath$\Delta_{\mbox{\bf\scriptsize all}}$ & \boldmath$\Delta_{\mbox{\bf\scriptsize 3,4}}$ & \boldmath$\Delta_{\mbox{\bf\scriptsize all}}$ & \boldmath$\Delta_{\mbox{\bf\scriptsize 3,4}}$ & \boldmath$\Delta_{\mbox{\bf\scriptsize all}}$ & \boldmath$\Delta_{\mbox{\bf\scriptsize 3,4}}$ & \boldmath$\Delta_{\mbox{\bf\scriptsize all}}$ & \boldmath$\Delta_{\mbox{\bf\scriptsize 3,4}}$ & \boldmath$\Delta_{\mbox{\bf\scriptsize all}}$ & \boldmath$\Delta_{\mbox{\bf\scriptsize 3,4}}$ & \boldmath$\Delta_{\mbox{\bf\scriptsize all}}$ & \boldmath$\Delta_{\mbox{\bf\scriptsize 3,4}}$ \\
\hline
{\bf 1.1} & 5.1\% & 3.3\% & 1.4\% & 0.5\% & 4.7\% & 0.4\% & 0.8\% & 0.3\% & 1.2\% & 0.1\% & 1.3\% & 1.0\% \\
{\bf 1.2} & 31.9\% & 31.9\% & 2.2\% & 0.3\% & 1.9\% & 0.9\% & 1.2\% & 0.1\% & 0.5\% & 0.3\% & 1.0\% & 0.8\% \\
{\bf 1.3} & 5.3\% & 3.0\% & 5.2\% & 0.4\% & 5.0\% & 1.9\% & 1.9\% & 0.3\% & 2.4\% & 0.8\% & 5.8\% & 4.6\% \\
{\bf 1.4} & 9.3\% & 9.0\% & 1.4\% & 0.5\% & 8.3\% & 2.9\% & 1.2\% & 0.5\% & 0.2\% & 0.0\% & 5.1\% & 1.7\% \\
{\bf 1.5} & 2.7\% & 2.7\% & 5.8\% & 3.4\% & 4.5\% & 3.0\% & 2.3\% & 1.3\% & 0.9\% & 0.6\% & 1.2\% & 0.7\% \\
{\bf 1.6} & 10.9\% & 10.3\% & 2.3\% & 0.3\% & 2.2\% & 0.2\% & 1.6\% & 0.2\% & 1.9\% & 0.2\% & 6.8\% & 0.5\% \\
{\bf 1.7} & 8.6\% & 1.7\% & 4.4\% & 1.0\% & 6.3\% & 0.9\% & 0.7\% & 0.3\% & 1.3\% & 0.2\% & 1.3\% & 0.9\% \\
\hline                
\end{tabular}
\label{tab:models_in_msa}
\end{center}
\end{table*}

\subsection{Global parameters and evolutionary sequences}

  For each case in Table~\ref{tab:modspec}, we have compared the global parameters calculated by the different codes: age, radius, luminosity, effective temperature, central temperature and density.
  The results are listed in Table~\ref{tab:models_in_msa} which include the spread in the parameters $(x_{i})$ as defined by:
  $$
  \Delta_{\beta} \equiv
     2\; \frac{\max(x_{i})-\min(x_{i})}{\max(x_{i})+\min(x_{i})}
  $$
  where the letter $\beta$ indicates which set of codes has been used to calculate the model: $\beta{=}$all means that values for all codes have been used, $\beta{=}4$ is only based on values for codes (A, C, L, S) and $\beta{=}3$ on values for codes (A, C, L) or (C, L, S).
  The subsets of codes have been selected because these codes have used a very similar implementation of the physics. 
  As can be seen in these tables when all models are considered in the comparisons, differences $\Delta_{\beta}$ in the global parameters may reach several percent and even more for the age or the mass of the convective core (this latter is not given in the tables).
  This is mainly explained by the fact that the reference set of physics has not been fully implemented in all codes.
  For the set of codes that have followed more closely the physics specified, the differences are significantly smaller as illustrated by $\Delta_{\mbox{\scriptsize 4}}$ and $\Delta_{\mbox{\scriptsize 3}}$.
  The corresponding evolutionary tracks in the HR diagram plotted in Figures~\ref{fig:dif_hr_inmsa} and \ref{fig:dif_hr_offms} are fairly consistent as expected from the comparison of the global parameters of the target models.

  Several iterations have been made in the comparison process.
  This has allowed us to find some errors or inaccuracies in the evolution codes or in the model calculation process \citep[see][and references therein]{monteiro05a,lebreton05,CD05ab,montalban05a,montalban05b}.
  As a result, the relative differences between the models have decreased. 
  However, further work has to be done for the codes which still do not follow exactly the physics specifications decided for Task~1 (Section~\ref{sec:physics}) and for which the greatest differences with the other codes are found.

\subsection{Internal structure}

\begin{figure}[ht!]
\centering
\begin{tabular}{ccc}
\hspace{-8pt}\includegraphics[width=0.9\linewidth]{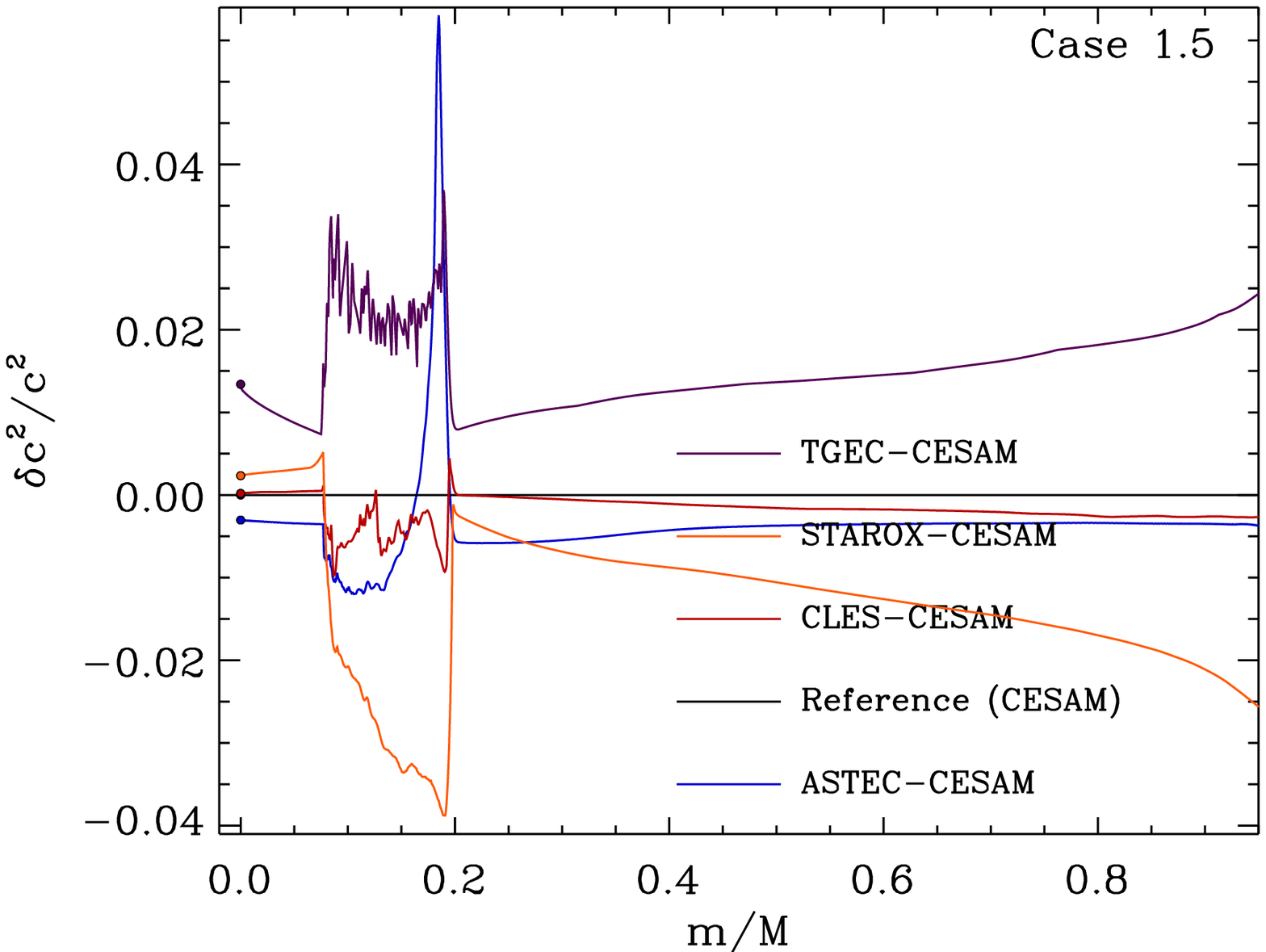} \\[-5pt]
\hspace{-8pt}\includegraphics[width=0.9\linewidth]{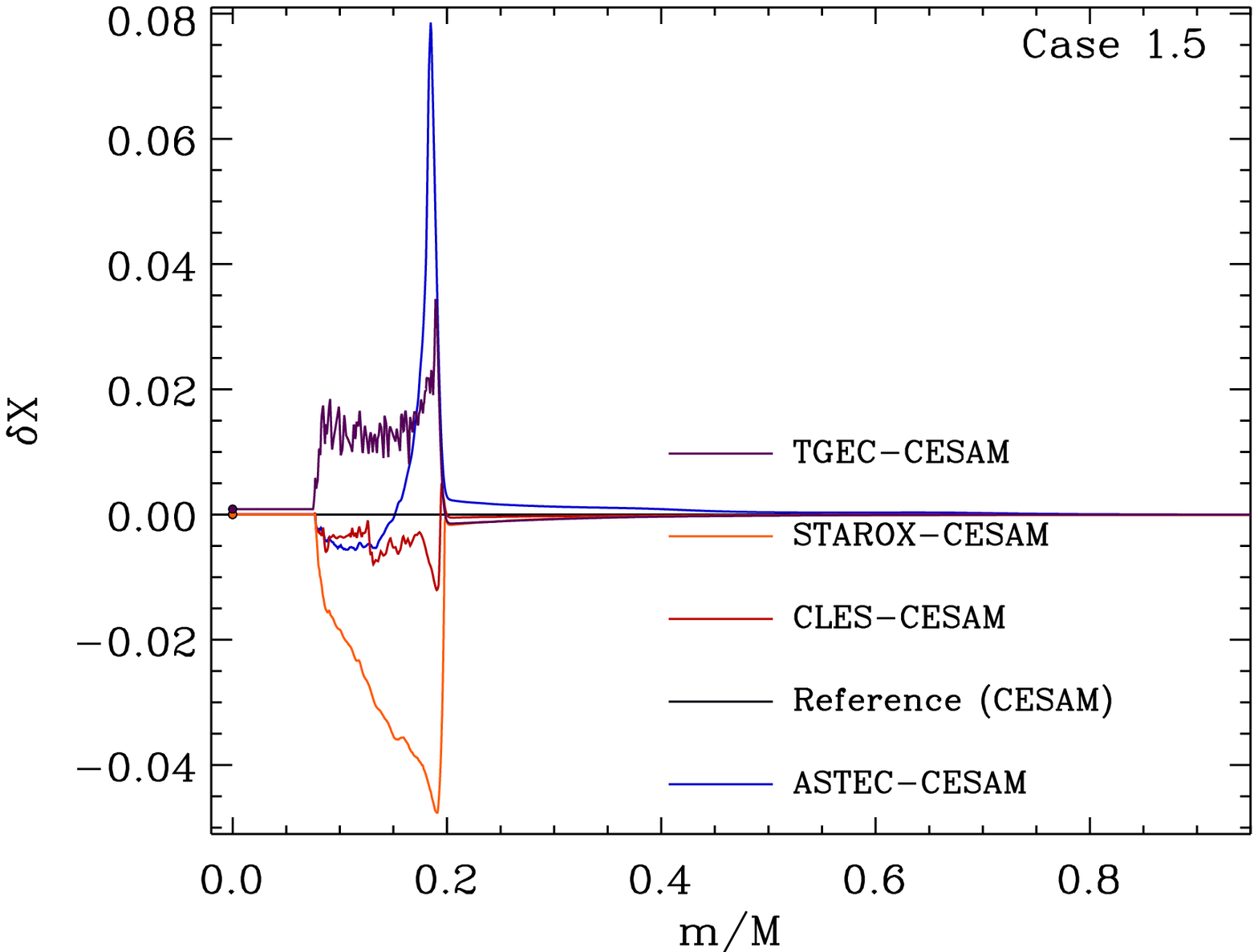}\\
\end{tabular}
\caption{
  Plot of the relative sound speed differences and hydrogen abundance differences at fixed relative mass between pairs of models corresponding to Case 1.5 (with overshooting).}
\label{fig:ss_xh_dif_a}
\end{figure}

\begin{figure*}[ht!]
\centering
\begin{tabular}{ccc}
\hspace{-8pt}\includegraphics[width=0.45\linewidth]{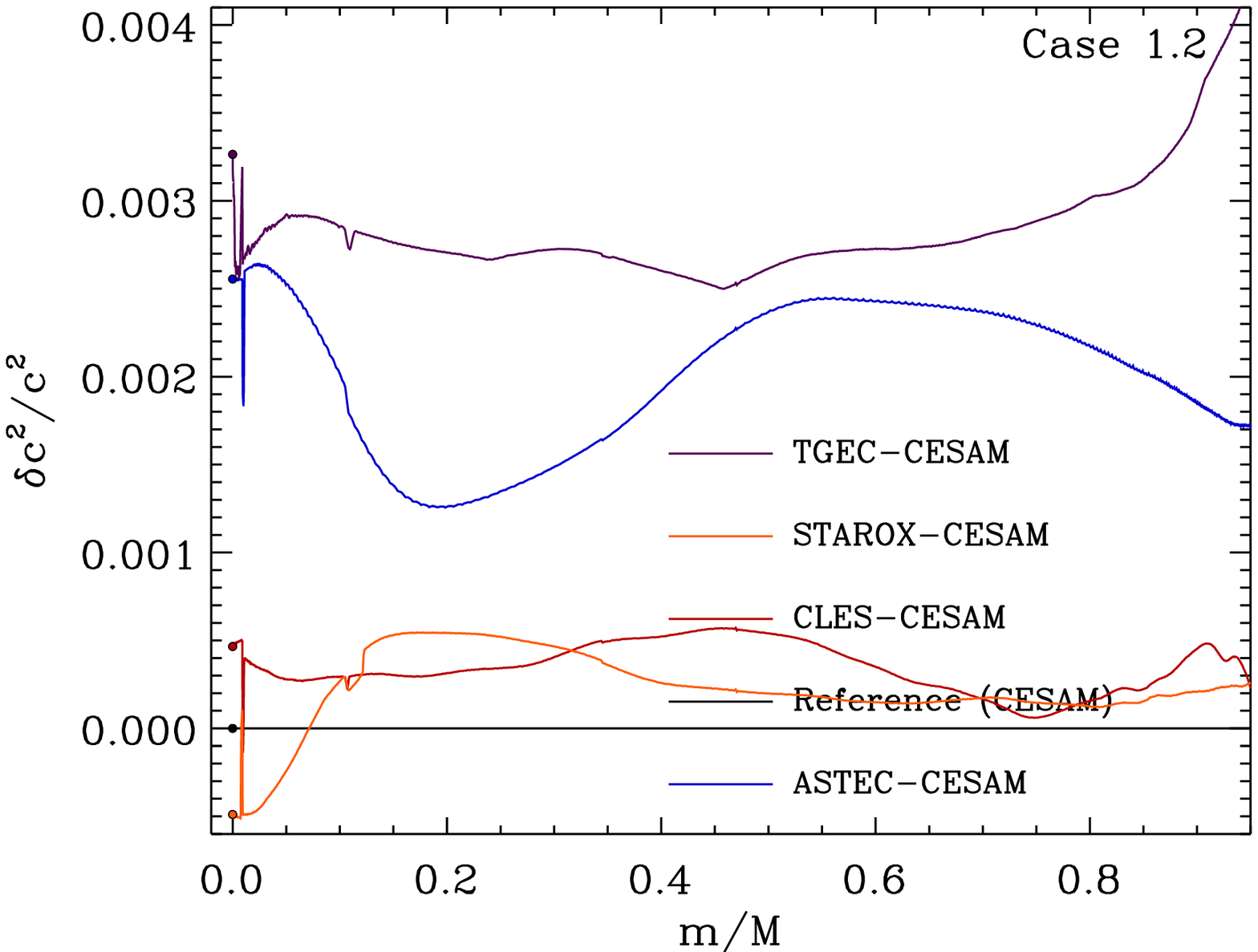} &
\hspace{-8pt}\includegraphics[width=0.45\linewidth]{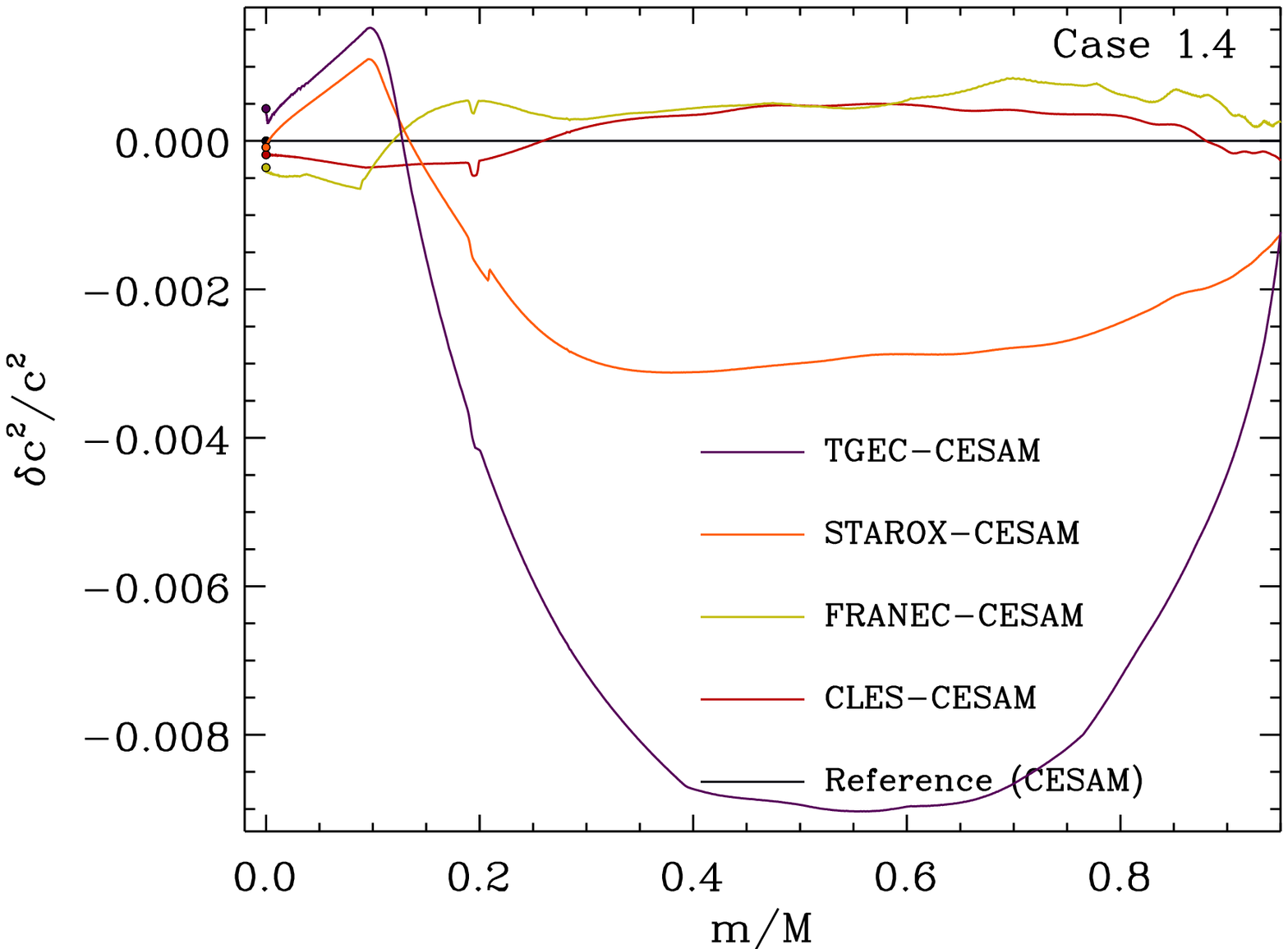} \\[-5pt]
\hspace{-8pt}\includegraphics[width=0.45\linewidth]{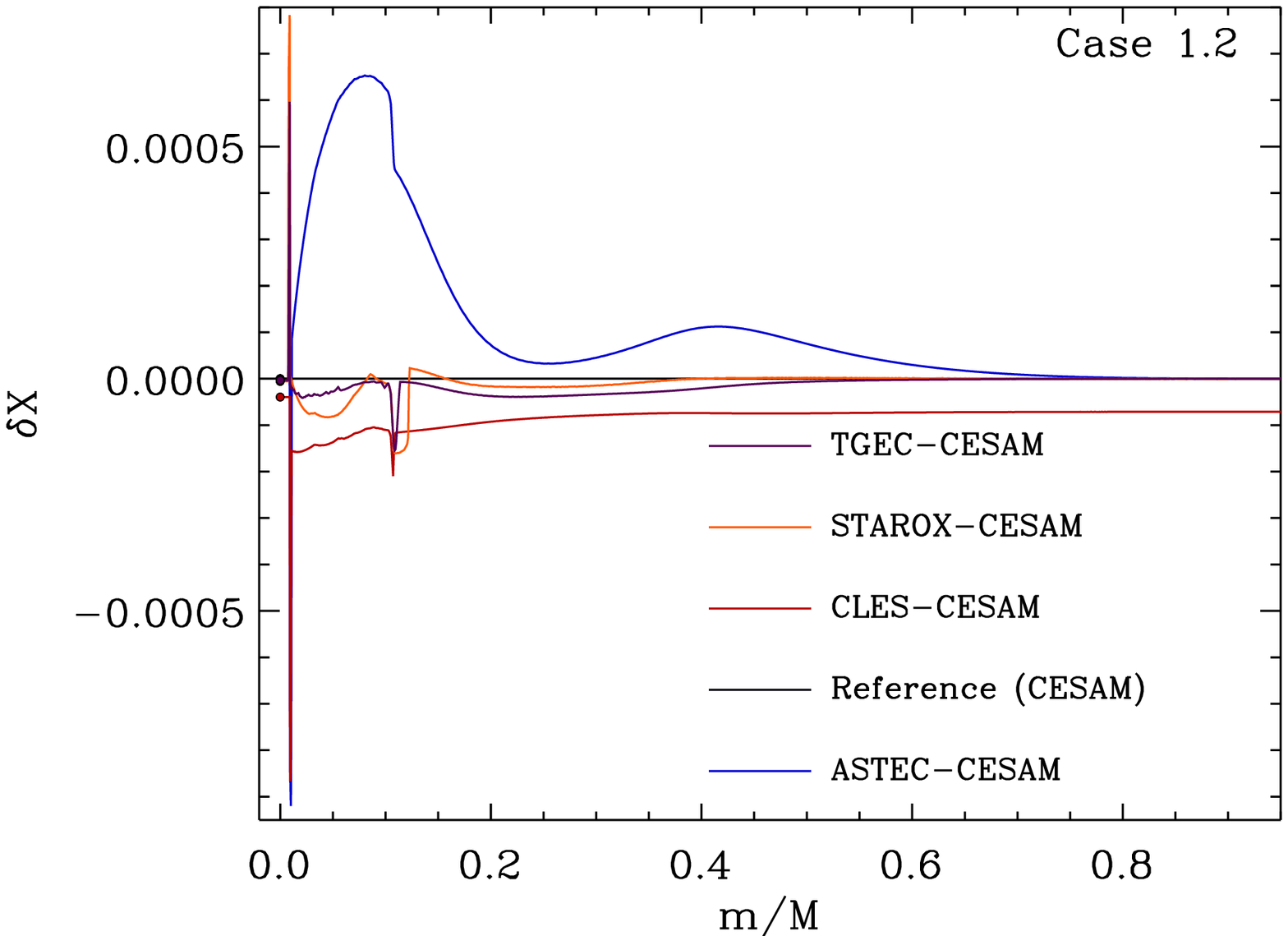} &
\hspace{-8pt}\includegraphics[width=0.45\linewidth]{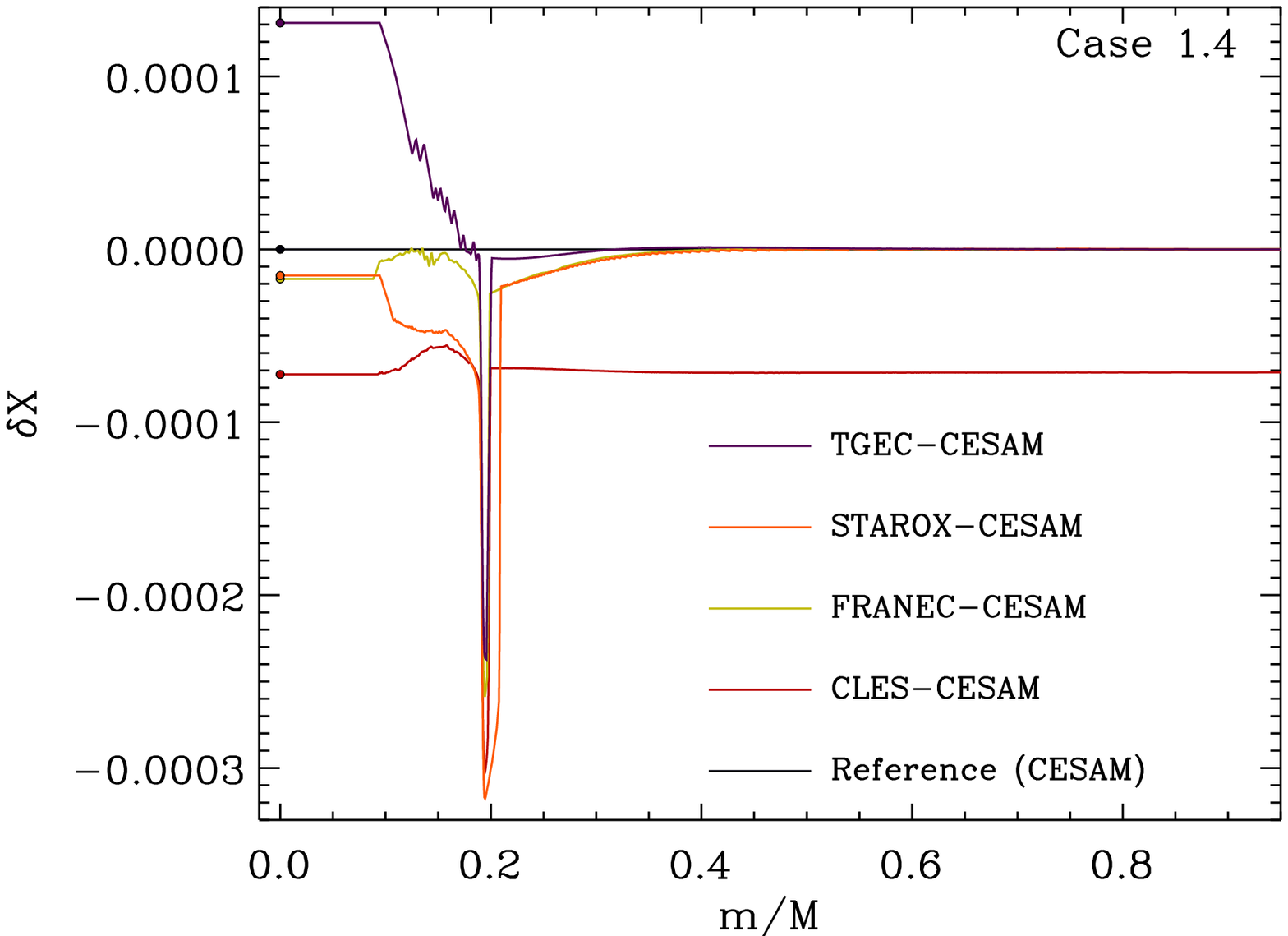}
\end{tabular}
\caption{
  Plot of the relative sound speed differences and hydrogen abundance differences at fixed relative mass between pairs of models corresponding to Case 1.2 (MS) and Case 1.4 (PMS).}
\label{fig:ss_xh_dif_b}
\end{figure*}

  We have compared the variations of the hydrogen abundance $X$ and of the square of the sound speed $c^2$ inside models calculated by different codes. 
  We have only used models whose global parameters are very similar in order to ensure that the differences are mainly determined by how each code calculates the evolution and the structure of the model and not by significant differences in the physics. 
  Some model differences in $X$ and $c^2$ are shown in Figures~\ref{fig:ss_xh_dif_a}--\ref{fig:ss_xh_dif_b} while Table~\ref{tab:models_dif} provides the spread of the model differences in the interior.
  The output from \cesam\ has been used as the reference for these differences.
  In this comparison we have excluded the outermost surface region ($m{>}0.95~M$) where we have found substantial differences between models; understanding the origin of these differences requires further work.

\begin{table}[ht!]
\begin{center}
\caption{
  Upper limits for the model differences in sound speed squared ($c^2$) and in hydrogen abundance ($X$) for the interior structure ($m/M{\le}0.95$) of the models being compared.
  The \cesam\ model is used as the reference for all cases while the last column indicates the codes used for the comparison.
  These values are a summary of the interior differences shown in  Figures~\ref{fig:ss_xh_dif_a}--\ref{fig:ss_xh_dif_b}.}        
\vspace{1em} 
\renewcommand{\arraystretch}{1.2}
\begin{tabular}[h]{lllc}
\hline \\[-9pt]
{\bf Case} & \boldmath$\displaystyle\max\left({\delta c^2 \over c^2}\right)$ & \boldmath$\max\left({\delta X} \right)$ & {\bf Codes} \\[9pt]
\hline
{\bf 1.1} & 0.0027 & 0.0022  & A, L, F, S \\
{\bf 1.2} & 0.0042 & 0.001   & A, L, S, T \\
{\bf 1.3} & 0.011  & 0.01    & A, L \\
{\bf 1.4} & 0.009  & 0.00032 & L, F, S, T \\
{\bf 1.5} & 0.06   & 0.08    & A, L, S, T \\
{\bf 1.6} & 0.01   & 0.012   & A, L, F, S \\
{\bf 1.7} & 0.018  & 0.02    & A, L, F, S \\
\hline
\end{tabular}
\label{tab:models_dif}
\end{center}
\end{table}

   One of the key aspects that dominates the model differences is the treatment of the borders of convective regions and how these change in time.
   This is clear in Case 1.5 (Figure~\ref{fig:ss_xh_dif_a}), corresponding to the 2~$M_\odot$ model on the TAMS with overshooting, large differences (up to 8\%) occur in the region of the convective core.
  Models differ in the treatment of the overshooting and of the convective boundaries and differences are spread in a mass range between $\sim$0.05 and 0.20~$m/M$ because the extent of the convective core decreases with evolution.
  
  The sound speed profile comparison also shows differences in the central regions as well as some differences in more external regions.
  These reflect differences in the treatment of the physics, in particular the equation of state and its implementation in the codes.
   The representation of the network of reactions is also the source of some discrepancies.
   Figure~\ref{fig:ss_xh_dif_b} illustrates the need to work further on how the different species are treated in the reactions network and a more consistent use of the equation of state tables, although the differences for these Cases are much smaller (below 0.1\% for $X$ and below 1\% for sound-speed squared).
   A factor in explaining the differences in the pre-main sequence phase is that different models are started with different initial conditions.
   
   These are some of the items that must be further studied in order to overcome part of the remaining model differences.
   Also, as demonstrated by \citet{montalban05a,montalban05b} from the comparison of the \cles\ and \cesam\ codes, differences can be reduced to a very small level (generally less than 0.5\%) when similar prescriptions are adopted for the input physics and their implementation.

\begin{figure}[ht!]
\centering
\begin{tabular}{cc}
\hspace{-8pt}\includegraphics[width=0.9\linewidth]{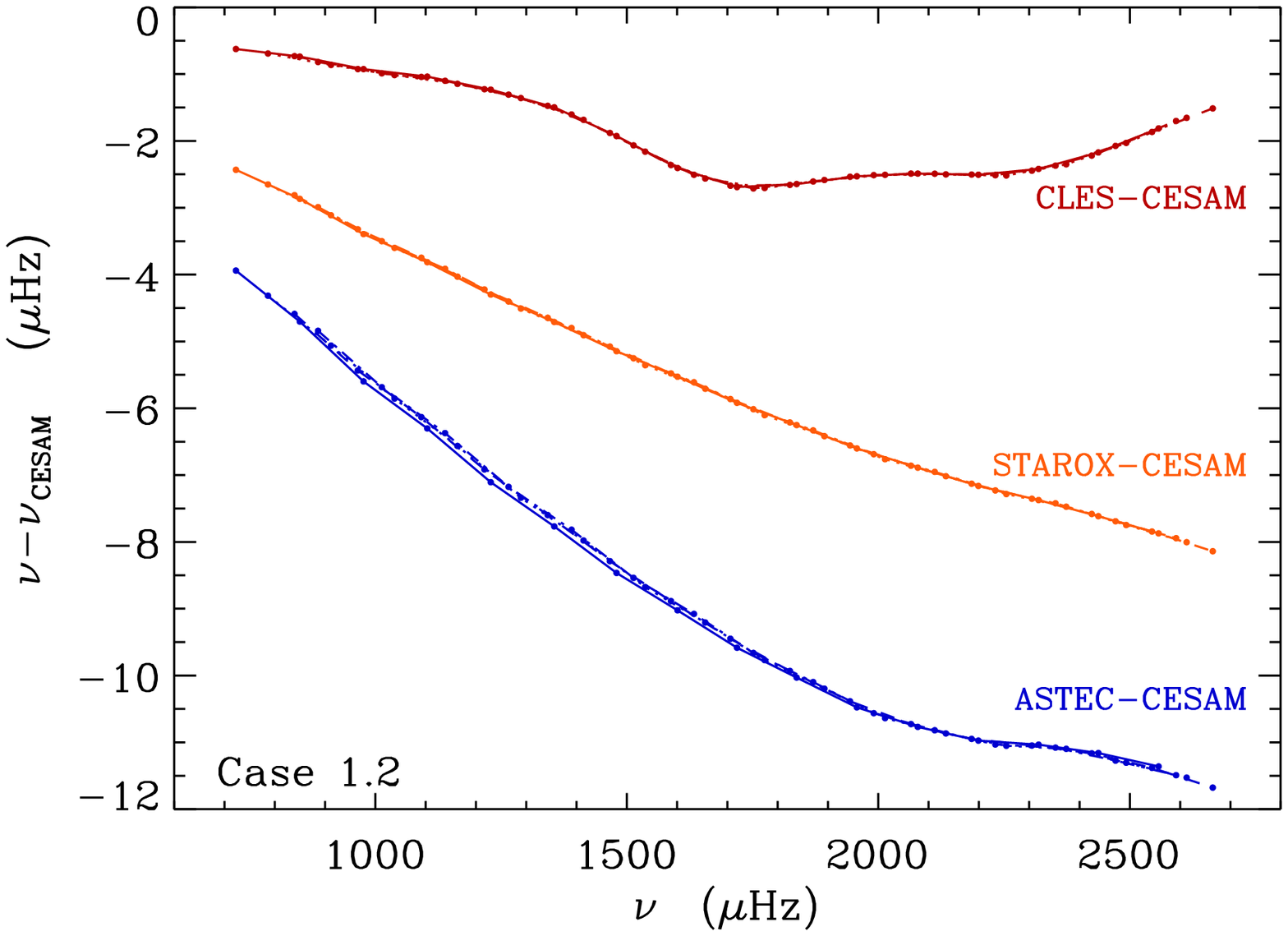} \\[-5pt]
\hspace{-8pt}\includegraphics[width=0.9\hsize]{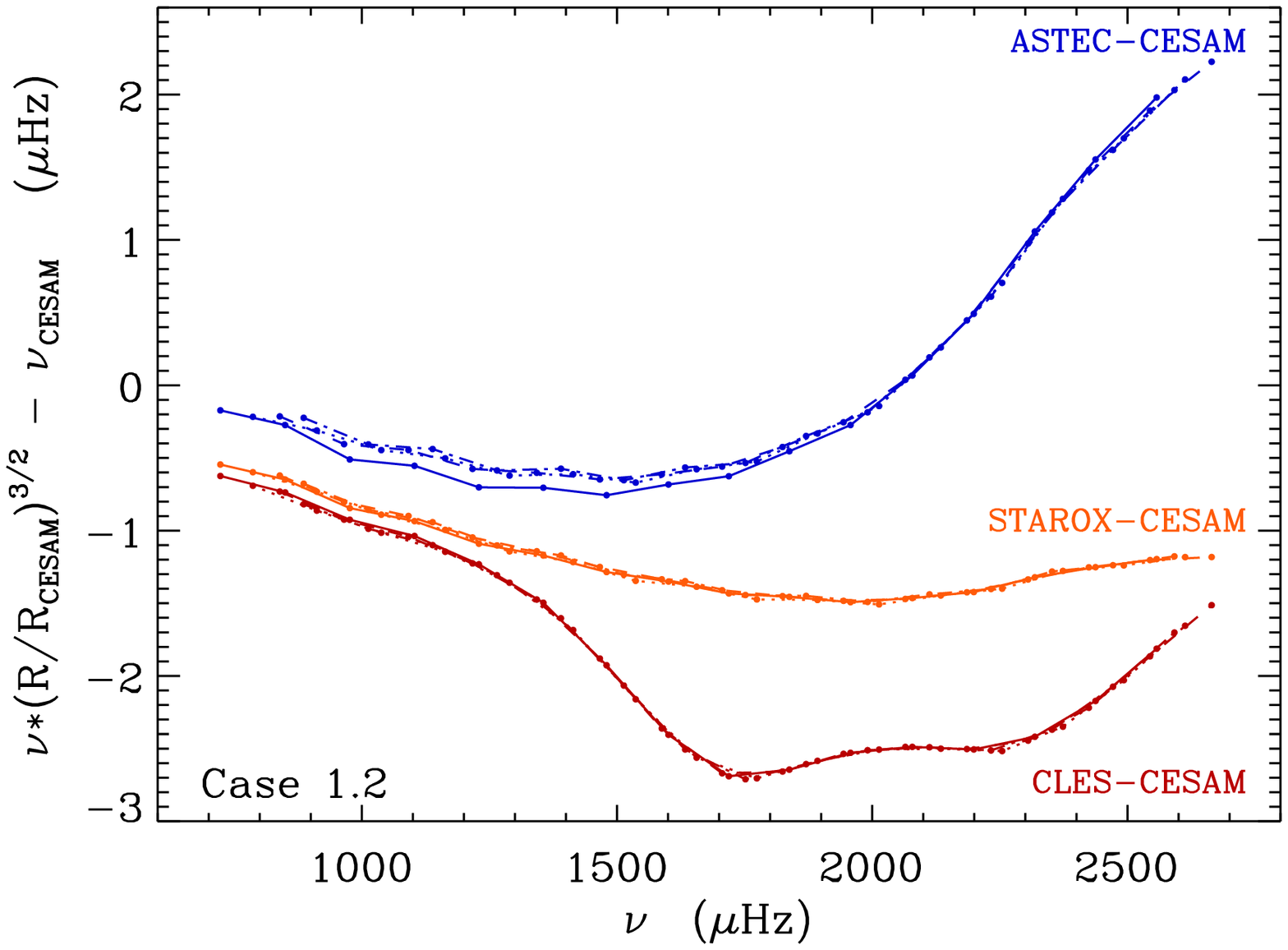}
\end{tabular}
\caption{
  Plot of the frequency differences, between models produced by different codes, for Case 1.2.
  The full line is for $\ell{=}0$, dotted line for $\ell{=}1$, dashed line for $\ell{=}2$ and dot-dashed line for $\ell{=}3$.
  Also shown are the frequency differences when the scaling due to the stellar radius ($R$) is removed.
  The remaining differences in the frequencies are mainly due to near surface effects.}
\label{fig:freq_dif}
\end{figure}

\subsection{Seismic properties}

\begin{figure}[ht!]
\centering
\begin{tabular}{cc}
\hspace{-8pt}\includegraphics[width=0.9\linewidth]{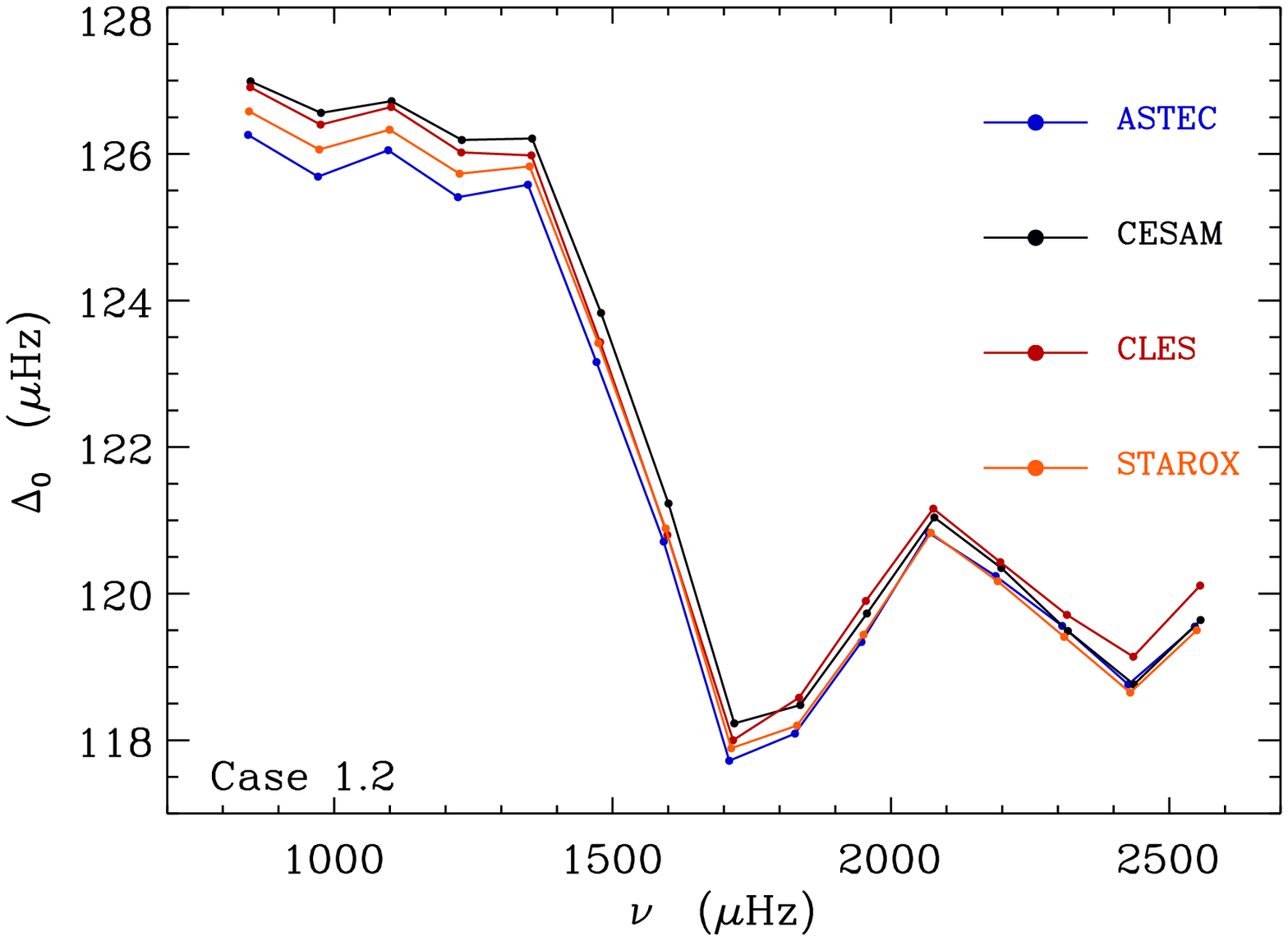} \\[-5pt]
\hspace{-8pt}\includegraphics[width=0.9\linewidth]{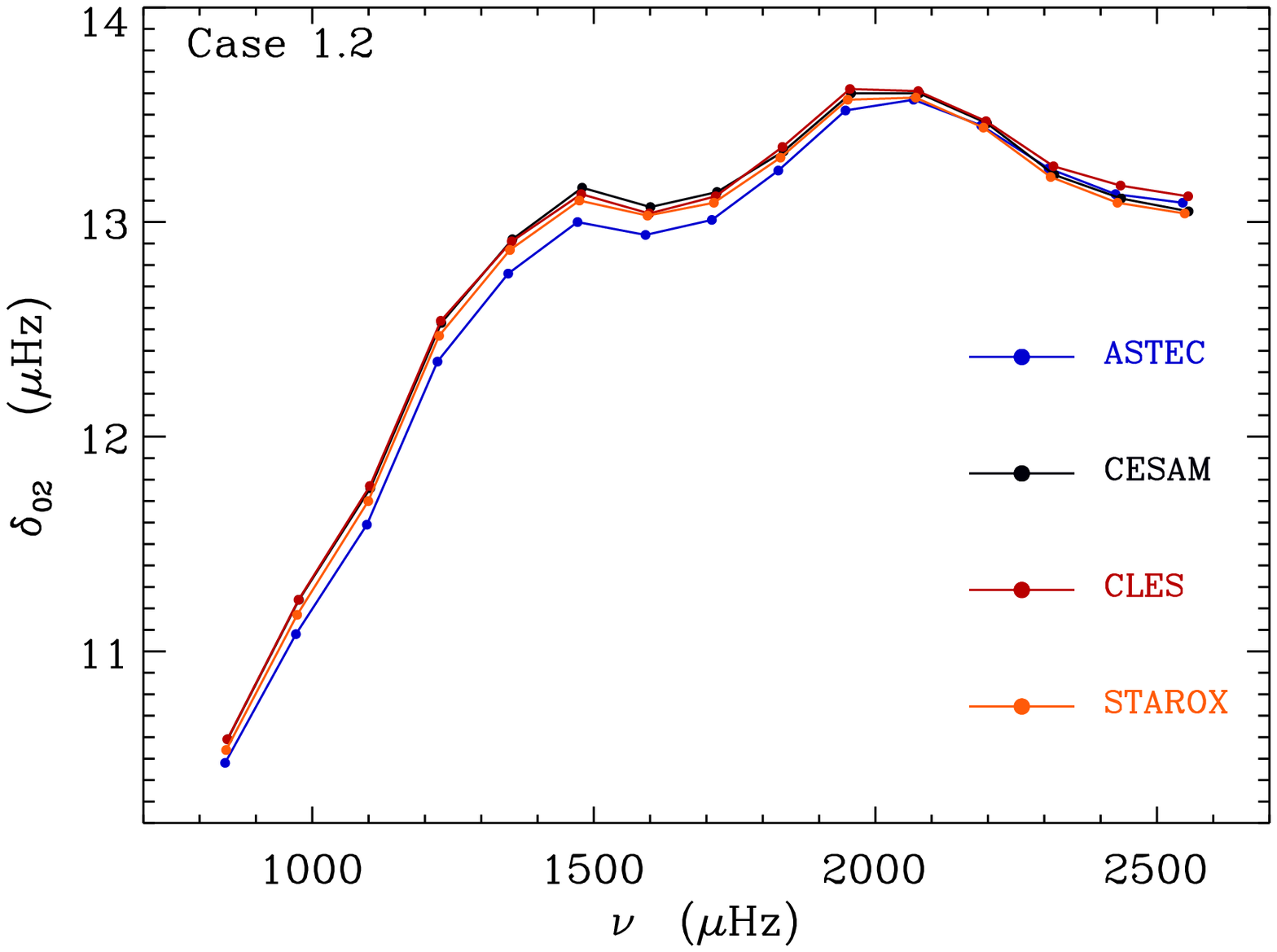}
\end{tabular}
\caption{
  Plot of the large ($\Delta_0$) frequency separation for Case 1.2.
  The small ($\delta_{02}$) frequency separations are also shown.}
\label{fig:sep_dif}
\end{figure}

  For the comparison of the seismic properties of the models we have used one seismic code \posc\ in order to calculate the frequencies of oscillations of all models for Case 1.2.
  We consider modes with degree $\ell{=}0,1,2,3$ and order $n{=}5{-}20$.
  In Figure~\ref{fig:freq_dif} the frequency differences are shown for these modes.
  After removing the scaling due to different stellar radius, the differences of the frequencies are dominated, as expected, by the near surface differences of the models.
  The component due to differences in the deep interior, as given by the differences between curves for different mode degrees, is small, being below the limit of the expected accuracy of the frequencies to be provided by \corot.
  For the same models in case 1.2, we also compared the frequency separations, i.e. the large frequency separation $\Delta_\ell{\equiv} \nu_{n{+}1,\ell} {-} \nu_{n,\ell}$ between the frequencies ($\nu_{n,\ell}$) for modes of degree $\ell{=}0$ and consecutive mode order $n$ and the small frequency separation $\delta_{\ell,\ell{+}2}{\equiv} \nu_{n,\ell} {-} \nu_{n{-}1,\ell{+}2}$ between the frequencies ($\nu_{n,\ell}$) for modes of degree $\ell{=}0$ and $\ell{=}2$ with mode orders $n$ and $n{-}1$ respectively (see Figure~\ref{fig:sep_dif}).
  The calculated separations are very similar, in accordance with the similarity of the global parameters of the models.

\section{Task~2: Frequency comparison}\label{sec:task2}

  In order to test the seismic codes being used we have started by comparing the frequencies of p-modes calculated using the same model (at exactly the same fixed mesh).
  For this exercise a 1.2~$M_\odot$ stellar model was used.
  The spectra considered in the comparison include frequencies of modes with degree $\ell{=}0,1,2$ and 3.
  In all comparisons we use \losc\ data as the reference.
  Some codes have used Richardson extrapolation and others have not.

\begin{figure}[ht!]
\centering
\hspace{-8pt}\includegraphics[width=0.9\linewidth]{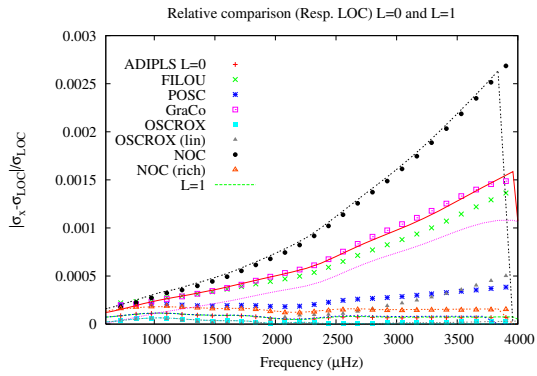} 
\caption{
  Relative frequency differences as a function of the frequency for modes with $\ell{=}0$ (dots) and $\ell{=}1$ (lines).}
\label{fig:freq}
\end{figure}

  Figure~\ref{fig:freq} shows the relative frequency differences for seismic data calculated with different codes (and different numerical procedures).
  Differences for $\ell{=}0$ modes present, in general, larger values than for $\ell{\ne}0$.
  The figure also confirms that the use of Richardson extrapolation reduces the discrepancies between codes.
  
\begin{figure}[ht!]
\centering
\begin{tabular}{cc}
\hspace{-8pt}\includegraphics[width=0.9\linewidth]{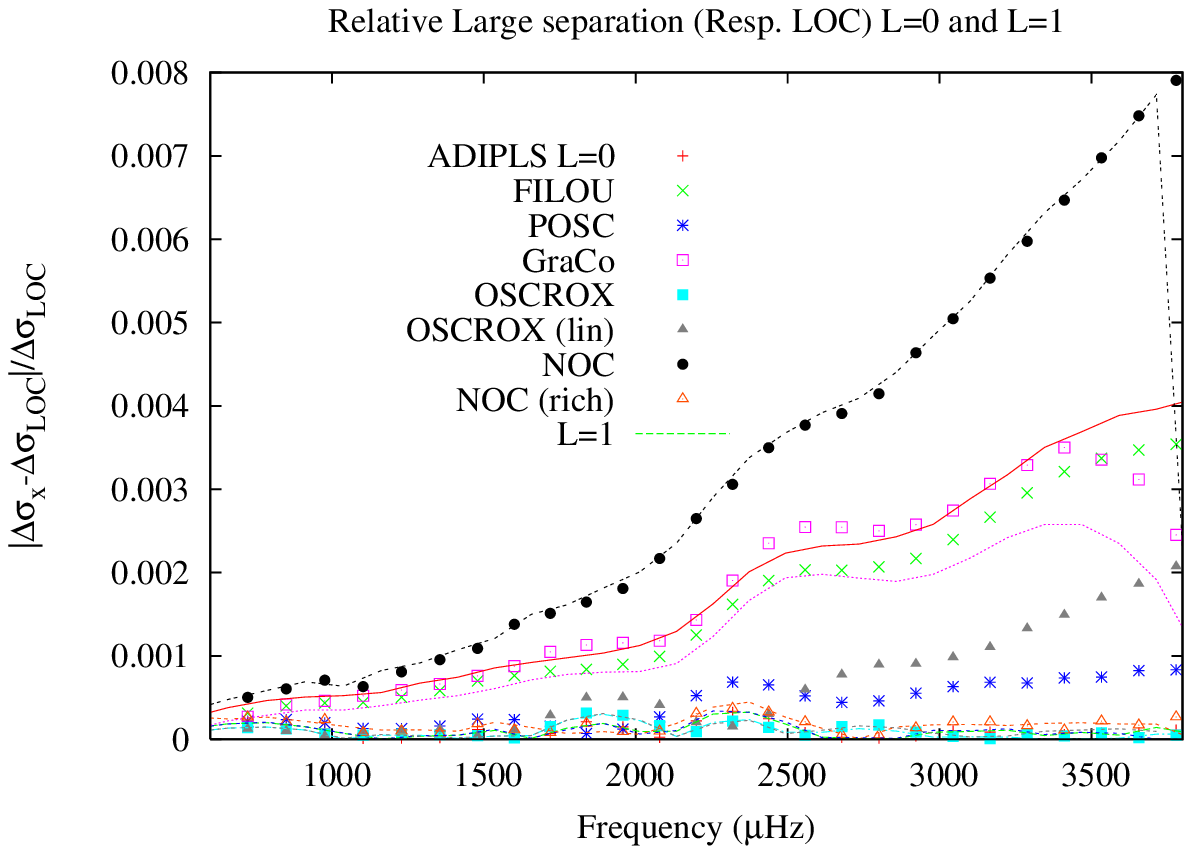}\\[0pt]
\hspace{-6pt}\includegraphics[width=0.9\linewidth]{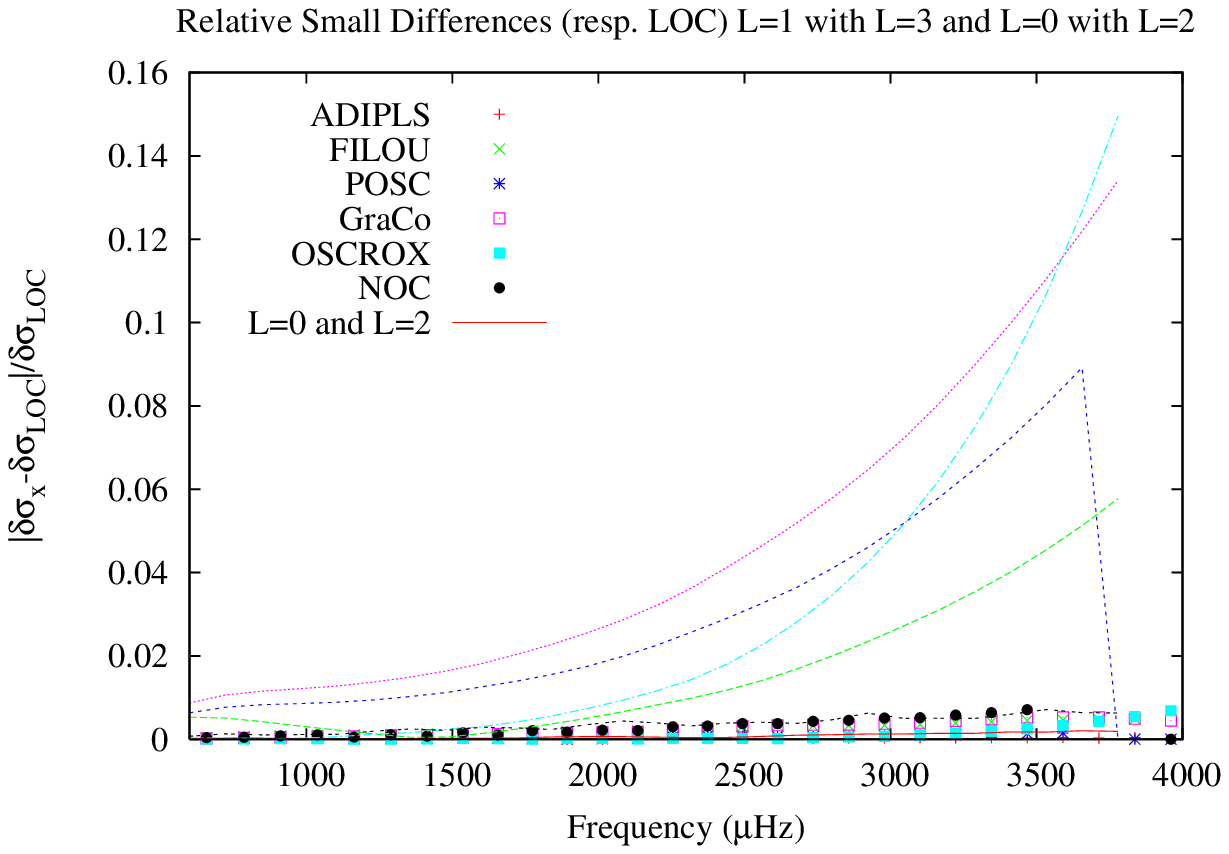}
\end{tabular}
\caption{
  Top: relative differences of the large separation as a function of the frequency for $\ell{=}0$ (dots) and $\ell{=}1$ (lines) modes.
  Bottom: relative differences of the small separation $\delta_{02}$ (lines) and $\delta_{13}$ (dots) as a function of the frequency for  modes.}
\label{fig:seps}
\end{figure}

  Figure~\ref{fig:seps} shows relative differences for the large separation (top) and small separation (bottom) for the same codes and spherical degrees as in Figure~\ref{fig:freq}.
  Codes using Richardson extrapolation present very similar results, while $\ell{=}0$ have larger differences than those coming from $\ell{=}1$.
  The relative differences of the small separation $\delta_{02}$ are one order of magnitude larger than $\delta_{13}$.

\begin{table}[ht!]
\begin{center}
\caption{
  Differences between frequencies and differences of the frequency separations calculated using the oscillation codes listed in Section~\ref{sec:tools} for a stellar model of 1.2~$M_\odot$ at the ZAMS.}
\vspace{1em} 
\renewcommand{\arraystretch}{1.2}
\begin{tabular}{|c|cc|cc|cc|}
\hline
 & \multicolumn{2}{c|}{\boldmath$\nu_{\ell,n}$} &
   \multicolumn{2}{c|}{\boldmath$\Delta_{\ell}$} & 
   \multicolumn{2}{c|}{\boldmath$\delta_{\ell,\ell{+}2}$} \\
 & \boldmath$\ell{=}0$ & \boldmath$\ell{=}1$ & \boldmath$\ell{=}0$ & \boldmath$\ell{=}1$ & \boldmath$\ell{=}0$ & \boldmath$\ell{=}1$\\
\hline
{\bf \boldmath$\mu$Hz} & 10 & 10 & 1 & 1 & 1.5 & 0.14 \\
{\bf \%} & 0.25 & 0.25 & 0.8 & 0.8 & 16 & 0.7 \\
\hline
\end{tabular}
\label{tab:freqs_dif}
\end{center}
\end{table}

  A summary of the results is listed in Table~\ref{tab:freqs_dif}.
  The major source of the differences is the use, or not, of Richardson extrapolation, but different codes react differently to the mesh as some are second order schemes while other used fourth order integration schemes \citep[see also][]{roxburgh05b}.
  A further step on the analysis on the sensitivity of the codes to the mesh of the models and the treatment of the boundary conditions is in progress \citep{moya05}.

\section{Reference grids}\label{sec:grids}

  In parallel to Task~1 and Task~2, grids of models have been especially calculated with the {\cesam} and {\cles} codes for masses in the range $0.8{-}10~M_\odot$ and chemical compositions $\mbox{[Fe/H]}{=}0.0$ and $-0.10$.
  These reference grids have already been used to locate \corot\ potential targets in the HR diagram and study some $\delta$ Scuti candidates for \corot\ \citep{2005AJ....129.2461P}.
  In addition oscillation frequencies for stellar models from the {\cesam} code have also been calculated either with {\posc} or {\adipls} codes.
  This material can be found on the ESTA Web site at {\tt\small http://www.astro.up.pt/corot/models/}.
  
\section{Conclusions and perspectives}\label{conclusion}

  The comparisons undertaken under Task~1 have shown that stellar models calculated by seven different codes are to first order consistent.
  Differences in the global parameters and in the internal hydrogen abundance and sound-speed profiles are generally in the range 1-10 percents except for the age and the extent of the convective core that may differ by 30\% or more.
  
  For codes that have followed precisely the reference set of physics defined for the exercise, the agreement is much better with differences of the order of 1 percent or less, which gives us confidence for future modelling of \corot\ targets by these codes. 
  Nevertheless, even if the reference set of physics is fixed, differences in the models remain due to different implementations of physics in the codes, different ways of interpolating in the tables, different choices of input parameters (isotopic ratios, atomic masses), etc.
  In a detailed comparison of the \cles\ and \cesam\ codes, \citet{montalban05a,montalban05b} have tried to identify these remaining inconsistencies showing that it should be possible to limit the differences to the order of 0.5\% or less.

  Concerning the global parameters, the source of the sometimes still large differences in age have to be examined which will imply the adoption of a common definition for the age zero.
  Concerning the seismic parameters, the differences in the total radius and in the structure of the surface layers are responsible for the main differences in the frequencies, thus requiring a more detailed comparison of these zones.
  
  Task~2 must be further developed before we can assess if the seismic codes are providing precise frequencies. 
  Further work on the mesh of the model, the numeric scheme used to solve the oscillations equations and the boundary conditions implemented is in progress.

  A detailed evaluation of the numeric precision and of the implementation of physics in the codes will continue to be necessary as the physics is improved and the methods and assumptions in the calculation become more complex.
  An example is the inclusion of diffusion to calculate the evolution -- how this is implemented requires a detailed study.
  Consequently, the work initiated by ESTA will proceed after \corot\ flies.
  As long as we need to calculate new stellar models with better physics, the need for a regular consistency analysis of the codes remains.


\end{document}